\documentclass[12pt,openbib]{article}
\usepackage{amsfonts}
\usepackage{amssymb}
\usepackage{amsmath}
\usepackage{graphicx}

 \newcommand{\R}{\mathbb{R}}

\newcommand{\dee}{\mathop{\! \,\mathrm{d} \!}\nolimits}
\newcommand{\comp}{\raisebox{0pt}{$\scriptstyle\circ \, $}}
\newcommand{\setrule}{\, \rule[-4pt]{.5pt}{13pt}\, }

\newcommand{\rowspace}{\rule{0pt}{16pt}}

\newcommand{\onehalf}{\mbox{$\frac{\scriptstyle 1}{\scriptstyle 2}\,$}}
\newcommand{\onequarter}{\mbox{$\frac{\scriptstyle 1}{\scriptstyle4}\,$}}

\begin{document}
\begin{center}{\Large \bf Monodromy in the resonant swing spring} \\
\mbox{} \vspace{.05in} \\
Holger Dullin\footnotemark ,
\, \, Andrea Giacobbe\footnotemark, \, \,
and
\addtocounter{footnote}{-1}\, \, Richard Cushman \footnotemark \end{center}
\addtocounter{footnote}{-1}\footnotetext{Mathematical Sciences,
Loughborough University, Loughborough, LE11 3TU  U.K.}
\addtocounter{footnote}{1}\footnotetext{Mathematics Institute,
University of Utrecht, 3508TA Utrecht, The Netherlands}
\begin{center}
\textbf{Abstract}
\end{center}
{\footnotesize This paper shows that an
integrable approximation of the spring pendulum, when tuned to be in
$1:1:2$ resonance, has monodromy. The stepwise precession angle of the
swing plane of the resonant spring pendulum is shown to be a
rotation number of the integrable approximation. Due to the monodromy, 
this rotation number is not a globally defined function of the integrals.
In fact at lowest order it is given by $\arg(a+ib)$
where $a$ and $b$ are functions of the integrals. The resonant swing spring 
is therefore a system where monodromy has easily observed physical 
consequences.}

\section{Introduction}

The spring pendulum or swing spring is one of 
the simplest possible mechanical  systems. It is a spring with one end 
fixed and at the other end is attached a mass that is acted on 
by a constant vertical gravitation field. The name swing spring 
comes from the fact that for appropriate initial conditions
the mass can either swing like a pendulum or bounce up and down like a 
spring. However, if the frequencies of the swinging and springing 
motion in linear approximation near the 
equilibrium are in resonance, these two types of motions are 
intricately intertwined. In particular, the following motion is 
easily observed. Starting with the weakly unstable vertical springing motion, 
the system evolves into a planar 
swinging motion. This swinging motion is also transient 
and the system returns to its original springing motion. This 
cycle then repeats. Lynch \cite{LynchE} observed that the 
orientation of the swing plane typically changes from one swinging 
phase to the next. 
Moreover, the angle between the swing planes of any two successive 
swinging phases
is constant. However, the angle between the swing planes depends on 
initial conditions. He called this phenomenon the stepwise precession of 
the swing plane of the swing spring. It is this phenomenon that we are going to 
explain both qualitatively and quantitatively. 

The swing spring has a long history that is well described in \cite{LynchH}.
The earliest comprehensive work on the planar spring pendulum is \cite{VG33}. 
This paper gives a classical treatment of the $1:2$ resonance using 
action angle variables. It is written in the spirit of the 
old quantum mechanics and was actually motivated
by the Fermi $1:2$ resonance in ${\mathrm{CO}}_2$.
The advent of modern quantum mechanics
seems to have made this type of analysis old-fashioned if not archaic. 
As Lynch \cite{LynchH} points out most of the previous work is only
concerned with the planar spring pendulum, so that the stepwise precession of the
swing plane cannot be found. Some progress on the three dimensional system 
was made in \cite{LynchE}. After that Holm and Lynch \cite{HL} found that the 
system can be approximated by the 3-waves system and derived 
a differential equation for the angle of the swing plane. 
This was done using ``pattern evocation in shape space'' \cite{MS}. 
We show that equation found by Holm and Lynch is exact and is nothing but the 
equation for the evolution of one the angles of the action-angle coordinates
of an integrable approximation to the resonant swing spring.
We then trace the origin of the stepwise precession to the existence
of Hamiltonian monodromy in this integrable approximation. 
Hamiltonian monodromy is an obstruction to the existence of global
action variables, which was first described in \cite{duistermaat80} (see also 
\cite{cushman-bates}). It generically appears around an equilibrium 
point of an integrable two degree of freedom Hamiltonian system 
whose linearization has a complex quartuple $\pm \alpha \pm i\, \beta $ of 
eigenvalues \cite{zung}. Such an equilibrium point is of focus-focus type.
The integrable approximation of the resonant swing spring has three
degrees of freedom. But after reduction of a symmetry, one obtains a 
two degree of freedom system with a focus-focus point as a relative equilibrium.  
Physically this corresponds to the pure springing motion of the system. 
For the purpose of the present paper the most important consequence 
of monodromy is that the rotation number of invariant tori, that is, the ratio
of their frequencies, near the singularity is not a single valued function.
Our main result is that the stepwise precession of the swing plane
is given by such a rotation number, which explicitly has the form
\begin{equation}
\Delta \vartheta =\arg( a + ib ) + \mathrm{O}(\sqrt{a^2+b^2}).
\label{eq-introone}
\end{equation}
Here $a$ and $b$ are simple function of the integrals of the system,
and $a,b \to 0$ near the equilibrium of the swing spring. The amazing
feature of (\ref{eq-introone}) is that it is not differentiable
at the origin. This means that no matter how small the initial perturbation
from the equilibrium is, one can always obtain all possible values
for $\Delta \vartheta$.
In general, such multivaluednes has been described for integrable foliations
near focus-focus singularitites by Vu Ngoc \cite{vungoc}.
In some sense our result is a special case of his. However, he did not 
study the influence of the Hamiltonian, but only the foliation. \medskip 

The paper is organised as follows. In section 2 we briefly recall the 
physics of the swing spring. Then we derive an integrable approximation 
which is valid near the resonant equilibrium point. This integrable system 
is then reduced to a one degree of freedom system in section 4. We then 
describe the 
geometry of the image of its energy momentum map and show that there is monodromy.
The dynamics of the swing angle $\vartheta$ is described in section 6 and
finally we obtain equation (\ref{eq-introone}) for the rotation number 
$\Delta \vartheta$, by approximating an elliptic integral.

\section{The physics of the swing spring} 

The spring pendulum is a point particle $r =(x,y,z)$ in ${\R }^3$ of mass $m$ 
attached to a spring which moves in a constant vertical gravitation field. 
Its potential energy is
\begin{equation}
\widetilde{V}(r) = mg \, z + \onehalf k
({\ell }_0 - \| r \|)^2,
\label{eq-s1one}
\end{equation}
where $\| r \| =\sqrt{x^2+y^2+z^2}$. The unstretched 
length of the
spring with spring constant $k$ is ${\ell }_0$ and $g$ is the constant of 
gravity. 
The motion of the swing spring is governed by Newton's equations
\begin{equation}
m\, \ddot{r} = -\mathrm{grad}\, \widetilde{V}(r).
\label{eq-s1two}
\end{equation}
The system is in equilibrium when
the forces of gravity and the spring balance, that is, when
\begin{equation}
\mathrm{grad}\, \widetilde{V}(r) = 0.
\label{eq-s1three}
\end{equation} 
Thus $x= y= 0$ and $z =-\ell $. The
equilibrium length $\ell $ of the spring is determined by
\begin{equation}
k(\ell - {\ell }_0) = mg.
\label{eq-s1four}
\end{equation}
\par
Measuring length in units of equilibrium length $\ell $, mass in units
of $m$ and time in units of $\sqrt{\frac{g}{\ell }}$ (which is the
period of small amplitude pendulum oscillations), we see that the
Hamiltonian of the swing spring on phase space $T^{\ast}{\R }^3$ with 
canonical coordinates $(x,y,z,p_x,p_y,p_z)$ is
\begin{displaymath}
\widetilde{H} = \onehalf (p^2_x+p^2_y+p^2_z) + \widetilde{U}(x,y,z),
\end{displaymath}
where
\begin{equation}
\widetilde{U}(x,y,z) = z +\onehalf {\nu }^2 {\Big (1-\frac{1}{{\nu }^2}
- \sqrt{x^2+y^2+z^2}\Big) }^2
\label{eq-s1five}
\end{equation}
and $\nu = \sqrt{\frac{k\ell }{mg}} =\sqrt{\frac{\ell }{\ell - {\ell }_0}}$. 
Since $\ell > {\ell }_0$, we
have $\nu >1$. This says that the frequency of the
spring oscillation is greater than the frequency of the small amplitude
pendulum oscillations. The reason for this is that the frequencies 
are \emph{not} independent because they are coupled by (\ref{eq-s1four}). 
Expanding the
potential $\widetilde{U}$ (\ref{eq-s1five}) about its stable equilibrium
$(0,0,-1)$ to cubic terms gives 
\begin{equation} \widetilde{U}(x,y,z) = 
\onehalf (x^2+y^2+z^2) - \mu (x^2+y^2)z, 
\label{eq-s1six} 
\end{equation}
where $z$ is the displacement from $-1$ and $\mu = \onehalf ({\nu }^2-1)
> 0$. Thus the cubic approximation to the Hamiltonian of the swing spring
in dimensionless variables with its equilibrium shifted to the origin is
the Hamiltonian
\begin{equation} 
H^{\nu } =\onehalf (p^2_x+p^2_y+p^2_z)
+ \onehalf (x^2+y^2+{\nu }^2\, z^2) - \mu (x^2+y^2)z
\label{eq-s1eight}
\end{equation}
on phase space $T^{\ast }{\R }^3$ with canonical
coordinates $Z  =(x,y,z,p_x,p_y,p_z)$. By rescaling the coordinates and
changing the time scale, we may consider $\mu $ to be a small parameter,
which measures the distance to the origin.
\par
The swing spring Hamiltonian
undergoes a $1:1:2$ resonance when $\nu =2$. In the
remainder of this paper we will study only this case. The Hamiltonian of
the cubic approximation of the resonant swing spring is
\begin{equation}
\mathcal{H} =  H_0 +V \, = \, \onehalf
(p^2_x+p^2_y+p^2_z) + \onehalf (x^2+y^2+4\, z^2) - \mu (x^2+y^2)z \,.
\label{eq-s1eightstar}
\end{equation}

\section{An approximating integrable system}

In this section we find an integrable approximation to the resonant 
swing spring
$(\mathcal{H}, T^{\ast }{\R }^3, $ $\omega = \dee x \wedge \dee p_x + \dee y
\wedge \dee p_y + \dee z \wedge \dee p_z)$ by averaging over the flow
of the quadratic part of $\mathcal{H}$. Due to the resonance there will be 
secular
terms at cubic order. From general results it is known that a 
large measure of initial conditions of the averaged system stays close to the 
original solutions for long times, see e.g. \cite{DS3}.

Clearly the Hamiltonian $\mathcal{H}$ (\ref{eq-s1eightstar}) is invariant under
rotation about the $z$-axis
\begin{equation}
\varphi: (s,Z) \mapsto {\varphi}_s(Z)
= \mbox{{\tiny $ \left( \begin{array}{ccc|ccc} \cos s &
-\sin s & 0 & 0 & 0 & 0 \\
\sin s & \cos s & 0 & 0 & 0 & 0 \\
0 & 0 & 1 & 0 & 0 & 0 \\ \hline
0 & 0 & 0 & \cos s & -\sin s & 0 \\
0 & 0 & 0 & \sin s & \cos s & 0 \\
0 & 0 & 0 & 0 & 0 & 1 \end{array} \right) $}} Z,
\label{eq-s2nine}
\end{equation}
which has momentum
\begin{equation}
L = xp_y -yp_x.
\label{eq-s2ten}
\end{equation}
The quadratic terms $H_0$ of the Hamiltonian $\mathcal{H}$ (\ref{eq-s1eightstar})
are in $1:1:2$ resonance. The flow of the
Hamiltonian vector field $X_{H_0}$ corresponding to $H_0$ generates the
oscillator action
\begin{equation}
\psi :(t, Z) \mapsto {\psi }_t(Z) =
\mbox{{\tiny $ \left( \begin{array}{ccc|ccc} 
\cos t & 0 & 0 & \sin t & 0 & 0 \\
0  & \cos t & 0 & 0 & \sin t & 0 \\
0 & 0 & \cos 2t & 0 & 0 &   \onehalf \sin 2t\\ \hline
-\sin t & 0 & 0 & \cos t & 0 & 0 \\
0 & -\sin t & 0 & 0 & \cos t & 0 \\
0 & 0  & -2\sin 2t & 0 & 0 & \cos 2t
\end{array} \right) $}} Z.
\label{eq-s2eleven}
\end{equation}
\par
To make the oscillator action a symmetry, we average the Hamiltonian 
$\mathcal{H}$
(\ref{eq-s1eightstar}) over the integral curves of $X_{H_0}$. We obtain
\begin{eqnarray}
\overline{\mathcal{H}}(Z) & = & \frac{1}{2\pi } \int^{2\pi}_0 \mathcal{H}
\comp {\psi }_t(Z) \, \dee t \nonumber \\
& = & H_0 -\frac{\mu }{8}\left[ (xp_x+yp_y)p_z +(x^2+y^2)z -
(p^2_x+p^2_y)z \right].
\label{eq-s2twelve}
\end{eqnarray}
By construction the Hamiltonian $\overline{\mathcal{H}}$ is invariant under the
oscillator action (\ref{eq-s2eleven}). Therefore $\{ \overline{\mathcal{H}}, 
H_0\} =0$, where
$\{,\}$ is the standard Poisson bracket corresponding to $\omega$.
It is straightforward to check that $\overline{\mathcal{H}}$ is also
invariant under the axial action (\ref{eq-s2nine}) and hence
$ \{ \overline{\mathcal{H}}, L \} =0$. Since the axial and oscillator actions 
commute, it follows that $\{ H_0, L \} =0$. Consequently, the system
$(\overline{\mathcal{H}}, H_0, L, T^{\ast }{\R }^3, \omega )$ is Liouville
integrable.
\par
For later purposes it is useful to study a different,
but equivalent Liouville integrable system. This equivalent system is obtained 
by a 
linear symplectic transformation that diagonalizes both, $H_0$ and $L$.
Consider the invertible linear map
\begin{equation}
\begin{array}{l}\Psi : T^{\ast}{\R }^3 \rightarrow T^{\ast }{\R }^3:
(\xi ,\eta ,\zeta, p_{\xi },p_{\eta }, p_{\zeta }) \mapsto 
(x,y,z,p_x,p_y,p_z) \, =  \\
\rowspace
\hspace{1.75in} = \frac{1}{\sqrt{2}}\left( p_{\eta }+\xi , p_{\xi } +
\eta , \zeta , p_{\xi }- \eta, p_{\eta }-\xi , 2p_{\zeta } \right) .
\end{array}
\label{eq-s2thirteen}
\end{equation}
Then $\Psi $ is symplectic, that is,
\[
{\Psi }^{\ast }(\omega ) = \dee
\xi \wedge \dee p_{\xi} +\dee \eta  \wedge \dee p_{\eta } +\dee \zeta
\wedge \dee p_{\zeta } \, = \, \widehat{\omega }.
\]
Moreover, $\Psi $ diagonalizes the momenta $L$ (\ref{eq-s2ten}) and
$H_0$ (\ref{eq-s1eightstar}), namely,
\begin{equation}
\widehat{L} = {\Psi }^{\ast }(L) \, = \, 
\onehalf (p^2_{\eta }+{\eta }^2) -\onehalf (p^2_{\xi }+ {\xi }^2)
\label{eq-s2fourteen}
\end{equation}and
\begin{equation}
{\widehat{H}}_0 = {\Psi }^{\ast }(H_0) \, = \, \onehalf
(p^2_{\xi }+{\xi }^2 + p^2_{\eta }+ {\eta }^2 +2(p^2_{\zeta }+ {\zeta }^2) ) 
\label{eq-s2fifteen}
\end{equation}
In new coordinates $\Xi
=(\xi ,\eta , \zeta , p_{\xi },p_{\eta }, p_{\zeta })$ on $(T^{\ast }{\R}^3,
\widehat{\omega })$ the averaged Hamiltonian $\overline{\mathcal{H}}$
(\ref{eq-s2twelve}) becomes
\begin{equation}
\widehat{H} = {\Psi }^{\ast }(\overline{\mathcal{H}}) \, = \,
{\widehat{H}}_0 + \lambda \, \left[ (\xi p_{\zeta } - \zeta p_{\xi } )\eta
-(\xi \zeta +p_{\xi }p_{\zeta}) p_{\eta } \right] ,
\label{eq-s2sixteen}
\end{equation}
where $\lambda = \frac{\mu \sqrt{2}}{8}$.
Hamilton's equations for the integral curves of
$X_{\widehat{H}}$ are
\begin{equation}
\begin{array}{rlcrl}
\dot{\xi } &= p_{\xi }-\lambda (\eta \zeta  +p_{\eta } p_{\zeta }) & \quad
 &
{\dot{p}}_{\xi } & = - \xi - \lambda (\eta p_{\zeta } - \zeta p_{\eta }) \\
\rowspace \dot{\eta } & = p_{\eta } - \lambda (\xi \zeta  + p_{\xi}p_{\zeta })
& \quad & {\dot{p}}_{\eta } & = -\eta - \lambda (\xi p_{\zeta }-
\zeta  p_{\xi }) \\
\rowspace \dot{\zeta } & = 2p_{\zeta } + \lambda
(\xi \eta - p_{\xi }p_{\eta }) & \quad & {\dot{p}}_{\zeta } & = -2 \zeta
+ \lambda (\xi p_{\eta } + \eta p_{\xi }) .
\end{array}
\label{3wave}
\end{equation}
\par
Introduce new momenta
\begin{equation}
J^1 = \onehalf ({\widehat{H}}_0 -\widehat{L}) \, = \,
\onehalf (p^2_{\xi }+{\xi }^2 + p^2_{\zeta }+ {\zeta }^2)
\label{eq-s2seventeen}
\end{equation}
and
\begin{equation}
J^2 = \onehalf ({\widehat{H}}_0 + \widehat{L})
\, = \, \onehalf (p^2_{\eta }+{\eta }^2 + p^2_{\zeta }+ {\zeta }^2),
\label{eq-s2eighteen}
\end{equation}
whose Hamiltonian vectors fields
$X_{J^1}$ and $X_{J^2}$ have flows giving the $S^1$-actions
\begin{equation}
{\varphi }^{J^1}_t:(t, \Xi ) \mapsto
\mbox{{\tiny $ \left( \begin{array}{ccc|ccc} 
\cos t & 0 & 0 & \sin t & 0 & 0 \\ 
0 & 1 & 0 & 0 & 0 & 0 \\
0 & 0 & \cos t & 0 & 0 & \sin t \\ \hline 
-\sin t & 0 & 0 & \cos t & 0 & 0 \\
0 & 0 & 0 & 0 & 1 & 0 \\ 
0 & 0 & -\sin t & 0 & 0 & \cos t \end{array}\right) $}} \Xi
\label{eq-s2nineteen} 
\end{equation} 
and
\begin{equation} {\varphi}^{J^2}_s:
(s, \Xi ) \mapsto \mbox{{\tiny $ \left(
\begin{array}{ccc|ccc} 
1 & 0 & 0 & 0 & 0 & 0 \\
0 & \cos s & 0 & 0 & \sin s & 0 \\
0 & 0 & \cos s & 0 & 0 & \sin s \\ \hline
0 & 0 & 0 & 1& 0 & 0 \\
0 & -\sin s & 0 & 0 & \cos s & 0 \\ 
0 & 0 & -\sin s & 0 & 0 &
\cos s \end{array}\right) $}} \Xi ,
\label{eq-s2twenty}
\end{equation}
respectively. Since $\{ J^1, J^2 \} =0$, these actions commute.
Moreover, they leave the Hamiltonian $\widehat{H}$ (\ref{eq-s2sixteen})
invariant. Thus $(\widehat{H}, J^1,J^2,T^{\ast }{\R }^3, \widehat{\omega})$
is a Liouville integrable system.
\par
Equation (\ref{3wave}) is a variant of the 3-wave system, see \cite{HL} 
and the references therein. The crucial difference of (\ref{3wave}) to the usual
presentation of the $3$-wave system is that we retain 
the linear terms. These terms usually are 
removed by the ansatz $\xi + ip_\xi = A \exp(it)$, etc., which 
leads to equations in the amplitudes $A$ etc., see \cite{HL}.
In our treatment we retain the linear terms, because they
determine the swing plane to lowest order. Keeping these terms
enables us to find the swing plane \emph{without} invoking the 
``pattern evocation in shape space'' hypothesis, as was done in \cite{HL}.

\section{Reduction to one degree of freedom}

In what follows we study the geometry of the energy momentum map
\begin{equation}
\mathcal{EM}:T^{\ast }{\R }^3 \rightarrow {\R }^3: \Xi
\mapsto (\widehat{H}(\Xi),J^1(\Xi ),J^2(\Xi))
\label{eq-s2twentyone}
\end{equation}
of the Liouville integrable system $(\widehat{H}, J^1,J^2,T^{\ast }{\R }^3,
\widehat{\omega })$. This energy momentum map is related to the energy
momentum map
\[
\widetilde{\mathcal{EM}}:T^{\ast }{\R}^3 \rightarrow {\R }^3:
Z \mapsto (\overline{\mathcal{H}}(Z),L(Z), H_0(Z))
\]
of the resonant swing spring system by $\mathcal{EM} = \Lambda
\comp \widetilde{\mathcal{EM}} \comp \Psi $. Here $\Psi $ is the linear
symplectic map (\ref{eq-s2thirteen}) and $\Lambda $ is the invertible
linear map { \onehalf \tiny $ \left( \begin{array}{crc} 2 & 0 &
0 \\ 0 & -1 & 1 \\ 0 & 1 & 1 \end{array} \right) $}. Thus the Liouville
integrable systems $(\overline{\mathcal{H}}, L,H_0,T^{\ast }{\R }^3, \omega )$ 
and
$(\widehat{H}, J^1,J^2,T^{\ast }{\R }^3, \widehat{\omega })$ are
equivalent. \medskip

\begin{figure}
\centerline{\includegraphics[width=16cm]{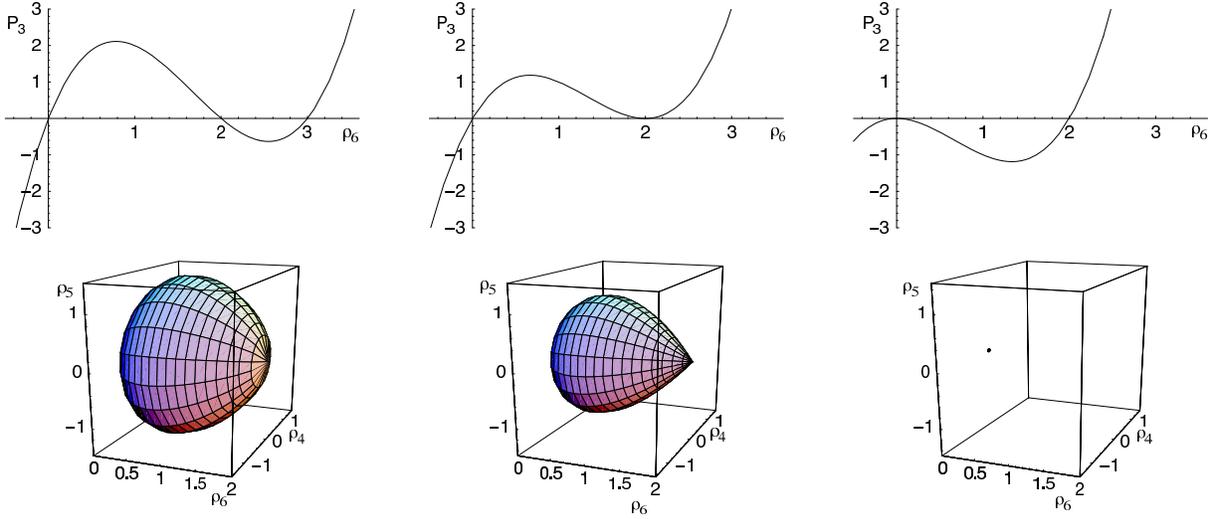}}
\caption{The polynomial $P_3$ (upper row) and the corresponding 
reduced spaces $P_{j_1,j_2}$ (lower row) illustrating the three types 
of reduced space using $j_1 = 1, j_2 = 3/2$ (left), $j_1 = j_2 = 1$ (middle), 
and $j_1 = 0, j_2 = 1$ (right).
\label{FigP}}
\end{figure}

The simplest way to reduce $(\widehat{H}, J^1,J^2,T^{\ast }{\R }^3, 
\widehat{\omega })$ is to consider the two-torus action
generated by the momenta $J^1$ and $J^2$. Concretely, define a $T^2$-action
\begin{equation}
\begin{array}{l}
\Phi : T^2 \times T^{\ast }{\R }^3
\rightarrow T^{\ast }{\R }^3: ((t,s), \Xi) \mapsto
{\varphi }^{J^1}_t \comp {\varphi }^{J^2}_s(\Xi) = \\
\rowspace \hspace{1in} = \, \mbox{{\tiny $\left( \begin{array}{ccc|ccc} 
\cos t & 0 & 0 & \sin t & 0 & 0 \\ 
0 & \cos s & 0 & 0 & \sin s & 0 \\ 
0 & 0 & \cos (t+s) & 0 & 0 & \sin (t+s) \\ \hline
-\sin t & 0 & 0 &\cos t & 0 & 0 \\ 
0 & -\sin s & 0 & 0 & \cos s & 0 \\
0 & 0& -\sin (t+s) & 0 & 0 & \cos (t+s)
\end{array} \right) $}} \Xi ,
\end{array}
\label{T2action}
\end{equation}
which comes from combining the $S^1$ actions (\ref{eq-s2nineteen}) and
(\ref{eq-s2twenty}). From their momentum maps we obtain the momentum
mapping
\begin{equation}
J:T^{\ast }{\R }^3 \rightarrow {\R }^2: \Xi
\mapsto (J^1(\Xi), J^2(\Xi))
\label{eq-s3thirtyseven}
\end{equation}
of $\Phi $. To reduce the integrable system $(\widehat{H}, J^1,J^2,
T^{\ast }{\R }^3,
\widehat{\omega })$ by the symmetry $\Phi$, we use invariant theory.
In this approach, the generators of the algebra of invariant
polynomials are used as new coordinates. The algebra of $T^2$-invariant 
polynomials is generated by
\begin{equation}
\rho_1 = p_\xi^2 + \xi^2, \quad
\rho_2 = p_\eta^2 + \eta^2, \quad
\rho_6 = p_\zeta^2 + \zeta^2 
\end{equation}
and
\begin{equation}
   \rho_4 = (\xi\eta - p_\xi p_\eta) p_\zeta - (\xi p_\eta + \eta 
p_\xi) \zeta, \qquad
   \rho_5 = (\xi\eta - p_\xi p_\eta) \zeta + (\xi p_\eta + \eta p_\xi) 
p_\zeta \,.
\end{equation}
The invariance of $\rho_4$ and $\rho_5$ is obvious if we write them as
\[
\rho_5 + i \rho_4 = (\xi - i p_\xi)( \eta - i p_\eta) (\zeta + i p_\zeta), 
\]
Note that the cubic part of $\widehat H$ is exactly $\lambda \, \rho_4$.
The invariants are subject to the relation
\begin{equation} 
\label{rhorel}
\rho_4^2 + \rho_5^2 = \rho_1\rho_2 \rho_6, \quad \rho_1 \ge 0, 
\rho_2 \ge 0, \rho_6 \ge 0\,.
\end{equation}
Therefore the space $P_{j_1,j_2} = J^{-1}(j_1,j_2)/T^2$ 
of orbits of $\Phi$ with
momentum $(j_1,j_2)$ is defined by (\ref{rhorel}) together with
\begin{equation}
    \rho_1 + \rho_6 = 2 j_1, \quad \rho_2 + \rho_6 = 2 j_2\,,
\end{equation}
which just expresses $J^1$ and $J^2$ in terms of invariants. 
Eliminating $\rho_1$ and $\rho_2$ gives
\begin{equation} 
\label{Gdef}
    G(\rho_4,\rho_5,\rho_6) = \rho_4^2 + \rho_5^2 - 
\rho_6(2j_1-\rho_6)(2j_2-\rho_6) = 0\,,
\end{equation}
where $0 \le \rho_6 \le \min(2j_1,2j_2)$. This defines the fully
reduced space $P_{j_1,j_2}$ as a submanifold of $\R^3$ with 
coordinates $(\rho_4,\rho_5,\rho_6)$.
Because of the restriction on $\rho_6$ and since
$\rho_4^2 + \rho_5^2$ is nonnegative, there are essentially three 
possibilities for the
geometry of $P_{j_1,j_2}$, see Figure~\ref{FigP}. These 
possibilities are determined
by the position of the roots of the polynomial
\begin{equation}
    P_3(\rho_6) = \rho_6 (2j_1 - \rho_6)(2j_2 - \rho_6)\,.
\label{P3def}
\end{equation}
\noindent 1. $j_1 \not  = j_2 \not = 0$.
The space $J^{-1}(j_1,j_2)$ is smooth and the action $\Phi$ is free.
Then the fully reduced space is either
\begin{displaymath}
P^{>}_{j_1,j_2}: {\rho }^2_4+{\rho }^2_5 = {\rho }_6(2j_1-{\rho }_6)
(2j_2 - {\rho }_6), \, \, \, j_1 > j_2 > 0, \, \, \,
0 \le {\rho }_6 \le 2 j_2,
\end{displaymath}
or
\begin{displaymath}
P^{<}_{j_1,j_2}: {\rho }^2_4+{\rho }^2_5 = {\rho}_6(2j_1-{\rho }_6)
(2j_2 - {\rho }_6), \, \, \, j_2 > j_1 > 0, \, \, \,
0 \le {\rho }_6 \le 2 j_1,
\end{displaymath}
which are each diffeomorphic to a \emph{smooth} $2$-sphere.
Every point on the fully reduced space reconstructs to a two-torus orbit of 
the free action $\Phi$ in phase space.
\medskip

\noindent 2. $j_1 = j_2 \not = 0$. The space $J^{-1}(j_1,j_1)$ is 
smooth and the action $\Phi$ has a fixed point $(\rho_4,\rho_5,\rho_6) = 
(0,0,2j_1)$ 
with isotropy group $S^1$. Then the fully reduced space
\begin{displaymath}
P_{j_1,j_1}: {\rho }^2_4+{\rho}^2_5 = {\rho }_6(2j_1-{\rho }_6)^2, 
\, \, \, j_1 > 0, \, \, \, 0 \le
{\rho }_6 \le 2 j_1,
\end{displaymath}
is a \emph{topological} $2$-sphere with one (conical) singular point 
$(0,0,2j_1)$. 
This singular point reconstructs to a pure springing motion on $\xi = 
\eta = p_{\xi }=
p_{\eta} =0$ and ${\zeta }^2 + p^2_{\zeta } = 2 j_1$.
The general point of the fully reduced space reconstructs to 
a two-torus in phase space which has angular momentum zero. This 
two-torus dynamically decomposes into an $S^1$ family of planar $S^1$ motions.
\medskip

\noindent 3. $j_1=0$ or $j_2=0$. The fully reduced spaces 
$P_{j_1,0}$, $j_1 >0$ and $P_{0,j_2}$, $j_2 >0$ each are a point 
${\rho }_4 = {\rho }_5 = {\rho }_6
=0$. In $T^{\ast }{\R }^3$ this reconstructs to a
pure swinging motion with nonzero angular momentum
on the circle ${\xi }^2 + p^2_{\xi }=2 j_1$,
$\eta = p_{\eta }=\zeta = p_{\zeta } =0$ (counterclockwise
in the $(x,y)$ plane projection)
or the circle ${\eta }^2 + p^2_{\eta }= 2 j_2$, $\xi = p_{\xi }=\zeta 
= p_{\zeta } =0$ (clockwise), respectively. If $j_1=j_2=0$, then $P_{0,0}$ is
the point ${\rho }_2 = {\rho }_5
= {\rho }_6 =0$, which reconstructs to the
equilibrium point $\xi = p_{\xi }=\eta = p_{\eta }=\zeta = p_{\zeta } 
=0$. \medskip

Using physical language, one would say that 
the projection of a generic motion of the resonant 
swing spring to the $(x,y)$ plane is elliptically, linearly, or 
circularly polarized,
corresponding to cases 1, 2, 3, respectively. \medskip

\begin{figure}
\centerline{\includegraphics[width=5cm]{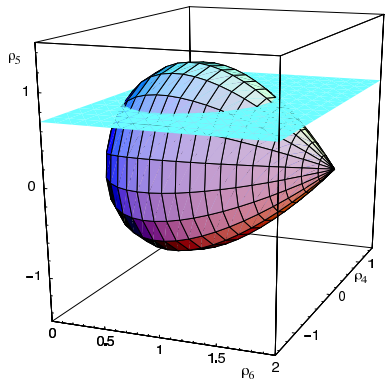}}
\caption{The intersection of the singular reduced space $P_{1,1}$ with the 
plane $h = const$ gives an integral curves of the reduced one degree of 
freedom system. 
\label{FigHG}}
\end{figure}

Because the Hamiltonian $\widehat{H}$ (\ref{eq-s2sixteen}) is
invariant under the $T^2$ action $\Phi $, it induces the fully reduced
Hamiltonian
\begin{equation}
H_{j_1,j_2}:P_{j_1,j_2} \subseteq {\R }^3
\rightarrow \R : ({\rho }_4, {\rho }_5,\rho_6) \mapsto j_1+j_2
+\lambda {\rho }_4.
\label{eq-s3thirtyninestar}
\end{equation}
The integral curves of the reduced system are now given by the intersection
of the reduced phase spaces $\{ G=0 \} $ and the planes $\{ H_{j_1,j_2}=h \}$, 
as illustrated in Figure~\ref{FigHG}. The fact that the Hamiltonian is a
linear function does \emph{not} imply that the system is a linear dynamical 
system, 
because the nonlinearity is contained in the Poisson bracket that
gives the dynamics on the reduced space.
A direct calculation shows that on $C^{\infty}({\R }^3)$, where ${\R }^3$ 
has coordinates $({\rho }_4, {\rho }_5,$ ${\rho }_6)$, the original 
canonical Poisson bracket on phase space $(T{\R }^3, \omega )$ 
induces a Poisson bracket $\{ \, , \, \} $ on ${\R }^6$ with 
coordinates $({\rho }_1, \ldots \, , {\rho }_6)$. Its nonzero 
brackets are given by
\begin{align}  \nonumber
\{\rho_1, \rho_4\} &= \{\rho_2, \rho_4\} = \{\rho_4, \rho_6\} 
= -2\rho_5, \\
\{\rho_1, \rho_5\} &= \{\rho_2, \rho_5\} = \{\rho_5, \rho_6\} = 
2\rho_4 , \\
\{\rho_4, \rho_5\} &= (\rho_1+\rho_2)\rho_6 -\rho_1\rho_2\,.  \nonumber
\end{align}
These brackets have the generators of the symmetry $\rho_1 + \rho_6$ and 
$\rho_2 + \rho_6$
as Casimirs, which we then use to eliminate $\rho_1$ and $\rho_2$. 
This shows that 
a Poisson bracket on the fully reduced space $P_{j_1,j_2}$ (considered as 
a subset of ${\R^3}$ with coordinates $(\rho_4,\rho_5,\rho_6)$) is 
\begin{eqnarray} \nonumber
\{ {\rho }_6, {\rho }_4 \} & =  & \frac{\partial G}{\partial {\rho }_5} 
\, = 2 {\rho }_5  \\
\{ {\rho }_4, {\rho }_5 \} & =  &\frac{\partial G}{\partial{\rho }_6} \, =                      {\rho}_6(2j_1+2j_2-{\rho }_6) - (2j_1-{\rho }_6)(2j_2-{\rho }_6)  \label{eq-s3forty} \\
\{ {\rho }_5, {\rho }_6\} & = & \frac{\partial G}{\partial {\rho }_4} \, = 
2 {\rho }_4, \nonumber
\end{eqnarray}
where $G$ is given in (\ref{Gdef}).

Hence the integral curves of the fully reduced Hamiltonian vector field
$X_{H_{j_1,j_2}}$ on $(P_{j_1,j_2}, \{ \, , \, \} )$ satisfy
\begin{equation}
\begin{array}{rl}
{\dot{\rho }}_4 & = \{ {\rho }_4, H_{j_1,j_2} \} \, = 0 \\
\rowspace {\dot{\rho }}_5 & = \{ {\rho }_5, H_{j_1,j_2} \} \,
= \lambda \, P'_3(\rho_6), \quad \mbox{see (\ref{P3def})} \\
\rowspace {\dot{\rho }}_6 & = \{ {\rho }_6,H_{j_1,j_2} \} \, = 
2\lambda \, {\rho }_5
\end{array}
\label{eq-s3fortystar}
\end{equation}
When $j_1=j_2 >0$, the $2j_1$-level set of the fully 
reduced Hamiltonian
$H_{j_1,j_1}$ is the intersection of the $2$-plane $\{ {\rho }_4 =0 \} $
with the fully reduced space $P_{j_1,j_1}$. This level set is an orbit
of $X_{H_{j_1,j_1}}$, which is homoclinic to the conical singular point
$(0,0,2j_1)$ of $P_{j_1,j_1}$. The conical singular
point reconstructs to the hyperbolic periodic orbit $\gamma $ of 
pure springing motion; while the homoclinic loop reconstructs 
to a $2$-torus bundle over each point of $H^{-1}_{j_1,j_1}(2j_1)
\setminus \{ (0,0,2j_1) \}$. Thus $H^{-1}_{j_1,j_1}(2j_1)$
reconstructs to the stable and unstable manifold in $T^{\ast }{\R }^3$ of
the hyperbolic periodic orbit $\gamma $. This invariant manifold is 
topologically the product of a once pinched $2$-torus and a circle. \medskip 

Describing the geometry of the $3$-tori and the dynamics of the swing
spring near this homoclinic invariant manifold is the main objective of
the remainder of this paper.

\section{The critical values of $\mathcal{EM}$}

We now determine the set of critical values of the energy momentum
mapping $\mathcal{EM}$ (\ref{eq-s2twentyone}) of the Liouville
integrable system $(\widehat{H}, J^1,J^2, T^{\ast }{\R }^3,
\widehat{\omega })$. The set of critical values of $\mathcal{EM}$
is called bifurcation diagram $\Sigma$. \medskip
\begin{figure}
\centerline{\includegraphics[width=5cm]{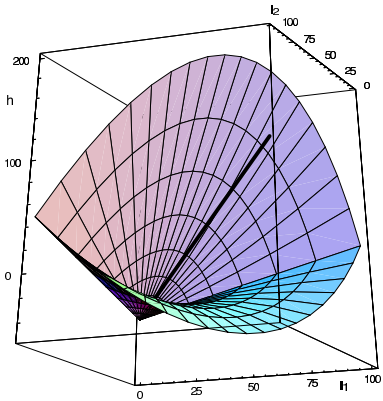}}
\caption{Set of critical values of the energy momentum map (bifurcation diagram).
It consists of two smooth patches intersecting
in two lines and a thread (thick black line) connected to the former only 
at the origin. The image of the momentum map contains the thread, hence
it is not simply connected. \label{FigSigma}}
\end{figure}
The critical values of $\mathcal{EM}$ can be are determined from
the equilibrium points of the reduced system.
Geometrically this occurs for energies $h$ for which the plane 
$\{ H_{j_1,j_2} =h \}$
intersects the fully reduced space $P_{j_1,j_2}$ in a (singular) point.
This occurs in different ways for the three types of reduced spaces.
In case 1, it occurs through a tangency of $\{ H_{j_1,j_2} =h\} $ 
with $P_{j_1,j_2}$.
In case 2, it is either again a tangency or the plane $\{ H_{j_1,j_2} =h \} $ 
contains the conical singular point $(0,0,2j_1)$. Hence 
$(j_1,j_2,h)=(j_1,j_1,2j_1)$.
In case 3, either $j_1$ or $j_2$ is zero, so that the critical
values are $(j_1,0,j_1)$ or $(0,j_2,j_2)$, respectively.

The points of tangency between $\{ H_{j_1,j_2} =h \} $ and $P_{j_1,j_2}$ 
(\ref{Gdef}) 
can be obtained using Lagrange multipliers. This leads to the condition 
that the polynomial
\begin{equation}
Q({\rho }_6) = -\frac{1}{\lambda^2}(j_1+j_2 -h)^2 + {\rho }_6(2j_1-{\rho }_6)
(2j_2-{\rho }_6)
\label{Qdef}
\end{equation}
has a multiple zero in $[0,\min (2j_1,2j_2)]$. When $j_1=j_2 \ge 0$
and $h =2j_1$, we see that $2j_1$ is a multiple root of $Q$ in $[0, 2j_1]$. 
\medskip 

The following argument determines a parametrization of the bifurcation 
diagram $\Sigma$. The bifurcation diagram $\Sigma$ is contained in the
set of $(h,j_1,j_2) \in {\R }^3$ where the polynomial $Q$ (\ref{Qdef}) has 
a multiple complex root. This set is the discriminant locus of the 
polynomial $Q$. 
If we restrict ourselves to the set of 
$(h,j_1,j_2) \in {\R }^3$ where the polynomial $Q$ (\ref{Qdef}) has a multiple
root in $[0, \min (2j_1, 2j_2)]$, then we obtain the restricted discriminant 
locus $\Delta $ of $Q$, which is equal the set of critical values $\Sigma$ of 
the 
energy momentum mapping $\mathcal{EM}$. When $(h,j_1,j_2) \in \Delta $, then 
$Q$ may be written as
\begin{eqnarray*}
Q({\rho}_6) & =
& {\rho }^3_6 -2(j_1+j_2){\rho }^2_6 +4j_1j_2 \, {\rho }_6 -
\frac{1}{{\lambda }^2}(j_1+j_2 -h)^2 \\
& = & ({\rho }_6 -s)^2({\rho }_6-t), \qquad s \in
[0, \min (2j_1, 2j_2)] \\
& = & {\rho }^3_6-(t+2s){\rho }^2_6 +(2st +s^2){\rho }_6 -s^2t.
\end{eqnarray*}
Comparing coefficients gives
\begin{equation}
\begin{array}
{rl}t+2s & = 2(j_1+j_2) \\
\rowspace 2st +s^2 & = 4j_1j_2 \\
\rowspace s^2t & =\frac{1}{{\lambda }^2}(j_1+j_2 -h)^2.
\end{array}
\label{eq-s4fortythree}
\end{equation}
Since $s \ge 0$, from the third equation in
(\ref{eq-s4fortythree}) it follows that $t \ge 0$. Taking the square
root of both sides of this equation gives
\begin{equation} j_1+j_2 -h =
{\varepsilon }_1 \lambda \, s\sqrt{t},
\label{eq-s4fortyfour}
\end{equation}
where ${\varepsilon }_1 = \pm $. Using the first two
equations in (\ref{eq-s4fortythree}), we see that
\begin{eqnarray}
4(j_1-j_2)^2 & = & 4(j_1+j_2)^2 -16j_1j_2 \nonumber \\
& = & (t+2s)^2 - 4(2st +s^2) \, = \, t(t-4s),
\label{eq-s4fortyfive}
\end{eqnarray}
which implies that
\begin{equation}
\mbox{either $t \ge 4s $ and $t \ge 0$, or $t=0$.}
\label{eq-s4fortyfivestar}
\end{equation}
When either of the conditions
(\ref{eq-s4fortyfivestar}) hold, we may take the square root of both
sides of (\ref{eq-s4fortyfive}). We then obtain
\begin{equation} j_1-j_2 =
\onehalf {\varepsilon }_2 \sqrt{t(t-4s)},
\label{eq-s4fortysix}
\end{equation} where ${\varepsilon }_2 = \pm $. Adding and subtracting
(\ref{eq-s4fortyfour}) from
\begin{equation}
j_1+j_2 = \onehalf (t+2s),
\label{eq-s4fortyseven}
\end{equation}
(see the first equation in (\ref{eq-s4fortythree})) gives
\begin{equation}
\begin{array}{rl}j_1 & =
j_1(s,t) \, = \, \onequarter (t+2s +{\varepsilon }_2\sqrt{t(t-4s)})\\
\rowspace j_2 & = j_2(s,t) \, = \, \onequarter (t+2s -
{\varepsilon}_2\sqrt{t(t-4s)}).
\end{array}
\label{eq-s4fortyeight}
\end{equation}
Subtracting (\ref{eq-s4fortyseven}) from (\ref{eq-s4fortyfour}) we
obtain
\begin{equation}
h = h(s,t) \, = \, \onehalf (t+2s) - {\varepsilon}_1 \lambda \,
s\sqrt{t}.
\label{eq-s4fortynine}
\end{equation}
\par
Consider the mapping
\begin{displaymath}
{\mathcal{P}}_{{\varepsilon }_1,
{\varepsilon}_2}: D \subseteq {\R }^2 \rightarrow {\R}^3: (s,t) \mapsto
(h(s,t),j_1(s,t),j_2(s,t)),
\end{displaymath}
where $D$ is the closed
subset of ${\R }^2$ which is the union of the intersection of the closed
half planes $\{ s \ge 0 \}$, $\{ t\ge 0\}$, $\{ t-4s \ge 0 \} $ and the
closed half line $L_{-} = \{ (s,0) \in {\R }^2 \setrule \, s \ge 0 \}$. 
For every choice of sign of ${\varepsilon }_1$ and ${\varepsilon }_2$,
the map ${\mathcal{P}}_{{\varepsilon }_1, {\varepsilon }_2 }$
parametrizes a patch of the bifurcation diagram $\Sigma $. All the 
patches together are shown in Figure~\ref{FigSigma}.

The image under ${\mathcal{P}}_{{\varepsilon }_1, {\varepsilon }_2 }$
of the closed half line $L_{-}$ is the thread $\mathcal{T} = \{ (2j_1, $ 
$j_1,j_1)$ $\in {\R }^3 \setrule \, j_1 \ge 0 \}$ , which is attached to the
$2$-dimensional pieces ${\mathcal{P}}_{{\varepsilon }_1, {\varepsilon
}_2 }(D\setminus L_{-})$ of $\Sigma $ only at the origin $(0,0,0)$.
Since the range $\mathcal{I}$ of the energy momentum mapping
$\mathcal{EM}$ (\ref{eq-s2twentyone}) is the closed region in ${\R }^3$
bounded by $\Sigma $ and containing the thread $\mathcal{T}$, we see
that the set of regular values $\mathcal{I}\setminus \mathcal{T}$ in the
image of $\mathcal{EM}$ is \emph{not} simply connected. Thus the
Liouville integrable system $(\widehat{H}, J^1, J^2, 
T^{\ast }{\R }^3,\widehat{\omega })$ possibly can have monodromy. The thread 
$\mathcal{T}$ represents the unstable springing motion
including its separatrix. The intersections of two patches, which 
form $2$-dimensional pieces in the boundary of the image $\mathcal{I}$, 
represent the periodic orbits of left or
right circular swinging motion. Any other point on the boundary
represents two-tori. Their special feature is that the instantaneous
ellipse in the $(x,y)$-plane formed by their projection has constant
excentricity, see section 8 below. All the remaining points in the 
image of $\mathcal{EM}$ represent generic motion on three-tori. \medskip 

For the geometrically inclined reader, we will show in sections 6 and 7 
that the integrable system $(\widehat{H}, J^1, J^2, T{\R }^3, 
\widehat{\omega })$ 
has monodromy. Analysts may skip these sections on the 
first reading and go directly to the derivation of the swing plane angle 
in section 8. This provides an analytic proof of the fact that the resonant 
swing spring has monodromy.

\section{A different reduction}

To uncover the geometry of the energy momentum map 
$\mathcal{EM}$ (\ref{eq-s2twentyone}), we reduce
the Liouville integrable system $(\widehat{H}, J^1,J^2,T^{\ast }{\R }^3,
\widehat{\omega })$ to a two degree of freedom Hamiltonian system by
removing the $S^1$-symmetry ${\varphi }^{J^1}_t$
(\ref{eq-s2nineteen}). As removing the symmetry induced by 
${\varphi }^{J^2}_s$ (\ref{eq-s2twenty}) leads to the one degree of
freedom system we have already discussed in section 4, we will 
omit this second reduction.  

To reduce the $S^1$-symmetry ${\varphi }^{J^1}_t$ we use invariant
theory. The algebra of polynomials on $T^{\ast }{\R}^3$, which is
invariant under this symmetry, is generated by
\begin{equation}
\begin{array}{rlcrlcrl}
{\sigma }_1 & = p^2_{\xi }+{\xi }^2 &\quad &
{\sigma }_3 & = \eta & \quad & {\sigma }_5 & = \xi p_{\zeta } - \zeta
p_{\xi } \\ \rowspace {\sigma }_2 & = p^2_{\zeta }+{\zeta }^2 & \quad &
{\sigma }_4 & = p_{\eta }&\quad & {\sigma }_6 & = \xi {\zeta }
+ p_{\xi } p_{\zeta }.
\end{array}
\label{eq-s3twentythree}
\end{equation}
These invariants are subject to the relation
\begin{equation}
{\sigma }^2_5 +{\sigma }^2_6 ={\sigma }_1 {\sigma }_2,
\, \, \, {\sigma }_1 \ge 0, \, \, {\sigma }_2 \ge 0,
\label{eq-s3twentyfour}
\end{equation}
which define the space $T^{\ast}{\R }^3/S^1$ of orbits of the
$S^1$-action ${\varphi }^{J^1}_t$. The
reduced space $P_{j_1} = (J^1)^{-1}(j_1)/S^1$ of orbits of
${\varphi}^{J^1}$ on the $j_1$-level set of the momentum $J^1$
(\ref{eq-s2seventeen}) is defined by (\ref{eq-s3twentyfour}) together with
\begin{equation}
\onehalf ({\sigma }_1 +{\sigma }_2) = j_1 \ge 0.
\label{eq-s3twentyfive}
\end{equation}
Eliminating ${\sigma }_1$ from
(\ref{eq-s3twentyfour}) using (\ref{eq-s3twentyfive}) gives
\begin{equation}
{\sigma }^2_5+{\sigma }^2_6 = {\sigma }_2(2j_1-{\sigma
}_2), \quad  0 \le {\sigma }_2 \le 2j_1,
\label{eq-s3twentysix}
\end{equation}
which defines $P_{j_1}$ as a submanifold of ${\R }^5$
(with coordinates $({\sigma }_2, \ldots , \, {\sigma }_6)$). Since the
Hamiltonian $\widehat{H}$ (\ref{eq-s2sixteen}) is invariant under
${\varphi }^{J^1}_t$, it induces the Hamiltonian
\begin{equation}
{\widehat{H}}_{j_1}:P_{j_1} \rightarrow \R :({\sigma }_2, {\sigma 
}_3, \ldots \, ,
{\sigma }_6) \mapsto j_1 +\onehalf ( {\sigma }_2 + {\sigma }^2_3 
+{\sigma }^2_4)
+\lambda ({\sigma }_3{\sigma }_5 - {\sigma }_4{\sigma }_6),
\label{eq-s3twentyseven}
\end{equation}
where $\lambda = \frac{\mu \sqrt{2}}{8}$.
\par
Consider ${\R}^6$ with coordinates $({\sigma }_1,{\sigma}_2,
\ldots , \, {\sigma }_6)$. On $C^{\infty}({\R}^6)$ there is a
Poisson bracket $\{\, , \, \}$, induced from the Poisson bracket on
the space of smooth functions on $(T^{\ast }{\R }^3, \widehat{\omega })$,
such that the nonzero brackets are
\begin{equation}
\begin{array}{rlcrlcrl}
\{ {\sigma }_5, {\sigma }_1 \} &
= 2{\sigma }_6 & \quad & \{ {\sigma }_6, {\sigma }_1\} & = -2{\sigma }_5
& \quad & \{ {\sigma }_5, {\sigma }_2 \} & = -2 {\sigma }_6 \\
\rowspace
\{ {\sigma }_6, {\sigma }_2\} & = 2{\sigma }_5 & \quad & \{ {\sigma }_4,
{\sigma }_3 \} & = -1 & \quad & \{ {\sigma }_6, {\sigma }_5 \} & =
{\sigma }_1 - {\sigma }_2.
\end{array}
\label{eq-s3twentyeight}
\end{equation}
Note that ${\sigma }_1+{\sigma}_2$ and ${\sigma }^2_5
+{\sigma }^2_6 -{\sigma }_1{\sigma }_2$ are Casimirs for the Poisson
algebra $(C^{\infty}({\R }^6), \{ \, , \, \}, \cdot )$. The reduced
Hamiltonian vector field $X_{{\widehat{H}}_{j_1}}$ on ${\R }^6$ has
integral curves which satisfy \begin{equation}
\begin{array}{rlcrl}
{\dot{\sigma }}_1 & = - 2\lambda ({\sigma }_3{\sigma }_6 + 
{\sigma }_4{\sigma }_5) &\quad & 
{\dot{\sigma }}_4 & = 
-\sigma_3 - \lambda {\sigma }_5 \\ 
\rowspace {\dot{\sigma }}_2 & = 2\lambda ({\sigma }_3{\sigma }_6 
+ {\sigma }_4{\sigma }_5) & \quad & 
 {\dot{\sigma }}_5 
& = -{\sigma }_6 +\lambda {\sigma }_4({\sigma }_1 - {\sigma }_2) \\ 
\rowspace  {\dot{\sigma }}_3 & = \sigma_4 - \lambda {\sigma }_6 & \quad &
{\dot{\sigma }}_6 & = {\sigma }_5 +  
\lambda {\sigma }_3({\sigma}_1-{\sigma }_2) .
\end{array}
\label{eq-s3twentynine}
\end{equation}

\section{Monodromy}

To show that the Liouville integrable system $(\widehat{H}, J^1, J^2,
T^{\ast }{\R }^3,\widehat{\omega })$ has monodromy, we fix $j_1>0$ and
reduce the $S^1$ action ${\varphi }^{J^1}_t$ (\ref{eq-s2nineteen}). We
obtain the two degree of freedom Liouville integrable system
$({\widehat{H}}_{j_1}, {\widehat{J}}^2, P_{j_1}, \{ \, , \, \} )$, where 
${\widehat{J}}^2 = \onehalf ({\sigma }_2 +{\sigma }^2_3 +{\sigma }^2_4)$ is 
induced from $J^2$ (\ref{eq-s2twenty}). We
will show that this system satisfies the hypotheses of the monodromy
theorem as stated in Cushman and Duistermaat, (see also Matveev \cite{matveev} 
and Zung \cite{zung}). Thus it has monodromy. This shows that the original
Liouville integrable system also has monodromy. \medskip

First, we verify that the reduced space $P_{j_1}$, $j_1 >0$ is a smooth
manifold. Since $P_{j_1} \subseteq {\R }^5$ is defined by
\begin{displaymath}
F(\widetilde{\sigma }) = {\sigma }^2_5 + {\sigma }^2_6 - 2j_1 
{\sigma }_2 + {\sigma }^2_2 =0, \, \, \, 0 \le {\sigma }_2 \le 2j_1,
\end{displaymath}
the point $\widetilde{\sigma }=({\sigma }_2,
\ldots , \, {\sigma }_6)$ is singular if
\begin{displaymath} (0,0,0,0,0)
= DF(\widetilde{\sigma }) \, = \, (2({\sigma }_2-j_1),0,0,2{\sigma }_5,
2{\sigma }_6),
\end{displaymath}
that is, ${\sigma }_2 = j_1$ and
${\sigma }_5={\sigma }_6 =0$. But $F(j_1, {\sigma }_3, {\sigma }_4, 0,0)
= -j^2_1 \ne 0$. Therefore $P_{j_1}$ has no singular points.
\par
Second, observe that the $S^1$ action on $P_{j_1}$ induced by 
the $S^1$ action ${\varphi }^{J^2}_s$ (\ref{eq-s2twenty}) is 
\begin{displaymath}
{\varphi }^{{\widehat{J}}^2}_s:(s, \sigma ) \mapsto 
\mbox{{\tiny $\left( \begin{array}{cc|cc|cc}
1 & 0 & 0 & 0 & 0 & 0 \\
0 & 1 & 0 & 0 & 0 & 0 \\ \hline 
0 & 0 & \cos s & -\sin s & 0 & 0 \\ 
0 & 0 & \sin s & \cos s & 0 & 0 \\ \hline
0 & 0 & 0 & 0 & \cos s & \sin s \\
0 & 0 & 0 & 0 & -\sin s & \cos s \end{array} \right) $}} \, \sigma . 
\end{displaymath}
The action ${\varphi }^{{\widehat{J}}^2}_s$ has a unique 
fixed point ${\widetilde{\sigma }}^0
=(2j_1,0,0,0,0)$ on $P_{j_1}\subseteq {\R }^5$. Since the reduced Hamiltonian
\begin{displaymath}
{\widehat{H}}_{j_1} = j_1 + \onehalf  ({\sigma }_2+{\sigma }^2_3+{\sigma }^2_4) 
+ \lambda ({\sigma }_3{\sigma }_5 -{\sigma }_4{\sigma }_6) 
\end{displaymath}
is invariant under ${\varphi }^{{\widehat{J}}^2}_s$, 
the point ${\widetilde{\sigma }}^0$ is
a critical point of ${\widehat{H}}_{j_1}$. Because $\{
{\widehat{H}}_{j_1}, {\widehat{J}}^2 \} =0$, the Hamiltonian vector
fields $X_{{\widehat{H}}_{j_1}}$ and $X_{{\widehat{J}}^2}$ commute. From
(\ref{eq-s3twentynine}) and the fact that ${\sigma }_1
+{\sigma }_2 = 2j_1$, we know that $X_{{\widehat{H}}_{j_1}}$ on ${\R}^5$ is
\begin{eqnarray*}
&& \mbox{{\footnotesize $ 2\lambda ({\sigma }_3{\sigma }_6 +
{\sigma }_4{\sigma }_5)
\frac{\partial}{\partial {\sigma }_2}
+\lambda {\sigma}_6 \frac{\partial}{\partial {\sigma }_3} +\lambda 
{\sigma }_5 
\frac{\partial}{\partial {\sigma }_4}$ }} \\
&&\hspace{.5in} \mbox{{\footnotesize $+({\sigma }_6 - 2\lambda {\sigma }_4
(j_1- {\sigma }_2))\frac{\partial}{\partial {\sigma }_5} - \, ({\sigma }_5 +  
2\lambda {\sigma }_3(j_1-{\sigma }_2))
\frac{\partial}{\partial {\sigma }_6}$.}}
\end{eqnarray*}
Clearly $X_{{\widehat{H}}_{j_1}}({\widetilde{\sigma}}^0) =0$.
Moreover, the linearization of $X_{{\widehat{H}}_{j_1}}$ on
the tangent space
\begin{displaymath} T_{{\widetilde{\sigma }}^0}P_{j_1}
= \ker DF({\widetilde{\sigma }}^0) \, = \, \ker (2j_1, 0,0,0,0) =
\mbox{{\footnotesize $\mathrm{span} \left\{ \frac{\partial}{\partial 
{\sigma }_3}, \,
\frac{\partial}{\partial {\sigma }_4}, \, \frac{\partial}{\partial
{\sigma }_5}, \, \frac{\partial}{\partial {\sigma }_6} \right\} $}}
\end{displaymath}
is
\begin{displaymath} Y =
DX_{{\widehat{H}}_{j_1}}({\widetilde{\sigma }}^0)|
T_{{\widetilde{\sigma }}^0}P_{j_1} =
\mbox{{\footnotesize $\left( \begin{array}{cccc} 0 & 0 & 0 & \lambda  \\
0 & 0 & \lambda & 0 \\
0& 2\lambda j_1 & 0 & 1 \\
2\lambda j_1 & 0 & -1 & 0 \end{array} \right) .$}}
\end{displaymath}
Since the characteristic polynomial of $Y$ is $(x-\lambda
\sqrt{2j_1})^2(x+\lambda \sqrt{2j_1})^2$, the equilibrium point
${\widetilde{\sigma }}^0$ is hyperbolic.

We now look at the energy momentum mapping
\begin{equation}
\begin{array}{rl}
{\mathcal{EM}}_{j_1}:P_{j_1} \subseteq {\R }^5
\rightarrow {\R }^2: \widetilde{\sigma } \mapsto &
({\widehat{H}}_{j_1}(\widetilde{\sigma }),
{\widehat{J}}^2(\widetilde{\sigma })) \, = \, \\
\rowspace &  \hspace{-1.75in} = \,
(j_1+\onehalf ({\sigma }_2 +{\sigma }^2_3+{\sigma }^2_4)
+\lambda ({\sigma }_3{\sigma }_5-{\sigma }_4{\sigma }_6),
\onehalf ({\sigma }_2 + {\sigma }^2_3+{\sigma }^2_4))
\end{array}
\label{eq-s5fiftyone}
\end{equation}
of the Liouville integrable system $({\widehat{H}}_{j_1},
{\widehat{J}}^2, P_{j_1}, \{ \, , \, \} )$ with fixed $j_1>0$. The fiber
${\mathcal{EM}}^{-1}_{j_1}(h,j_2)$ is compact. To see this note that
${\sigma }_2 \ge 0$ on $P_{j_1}$ and ${\sigma }_2 +{\sigma }^2_3 
+{\sigma }^2_4 =
j_2 \ge 0$. Therefore $0 \le {\sigma }_2 \le j_2$, $0 \le {\sigma }_3
\le \sqrt{j_2}$ and $0 \le {\sigma }_4 \le \sqrt{j_2}$. But
\begin{displaymath}
{\sigma }^2_5 +{\sigma }^2_6 = {\sigma }_2 (2j_1 -{\sigma }_2) ,
\quad 0 \le {\sigma }_2 \le 2j_1
\end{displaymath}
is the defining equation of $P_{j_1}$. Thus ${\sigma }_5$ and 
${\sigma }_6$ are
bounded. Hence ${\mathcal{EM}}^{-1}_{j_1}(h,$ $j_2)$ is compact. 
In other words, ${\mathcal{EM}}_{j_1}$ is a proper map. To show that
${\mathcal{EM}}^{-1}_{j_1}(h,j_2)$ is connected, note that it is the total
space of an $S^1$ bundle over the $h$-level set of the fully reduced
Hamiltonian $H_{j_1,j_2}$ (\ref{eq-s3thirtyninestar}) on the fully
reduced space $P_{j_1,j_2}$ (\ref{eq-s3thirtyseven}). Since this level
set is connected, when $(h,j_1,j_2)$ lies in the image of $\mathcal{EM}$
(\ref{eq-s2twentyone}) (and thus $(h,j_2)$ lies in the image of
${\mathcal{EM}}_{j_1}$ (\ref{eq-s5fiftyone})), it follows that
${\mathcal{EM}}^{-1}_{j_1}(h,j_2)$ is connected. At the critical value
$(2j_1,j_1)$ of ${\mathcal{EM}}_{j_1}$ we have already observed that,
reconstructing the critical $2j_1$-level set of the fully reduced
Hamiltonian $H_{j_1,j_1}$ on $P_{j_1,j_1}$, we obtain a once
pinched $2$-torus, which is the critical fiber
${\mathcal{EM}}^{-1}_{j_1}(2j_1,j_1)$. Thus we may apply the monodromy 
theorem. We find that over a loop $\Gamma $ in the set of regular values in
the image of ${\mathcal{EM}}_{j_1}$,
where $\Gamma $ encircles the critical value $(2j_1,j_1)$, the $2$-torus
bundle ${\mathcal{EM}}^{-1}_{j_1}(\Gamma )$ has monodromy {\tiny
$\left( \begin{array}{cc} 1 & 1 \\ 0 &1 \end{array} \right) $}.

\section{The swing plane angle}

In the original phase space $T^{\ast }{\R }^3$ with canonical
coordinates $(x,y,z,p_x,p_y,p_z)$ let us project the motion of the swing 
spring onto the $(x,y)$ plane of its configuration space ${\R }^3$.
Taking into account only the relevant quadratic terms
\begin{equation}
{{H}}^0_{xy} = \onehalf (p^2_x +p^2_y +x^2 + y^2),
\label{eq-s6fiftythree}
\end{equation}
of the Hamiltonian $\overline{H}$ (\ref{eq-s2sixteen}) of the resonant swing
spring, we obtain the Hamiltonian of a two degree of freedom harmonic 
oscillator. The projected motion of this oscillator is an ellipse $E$. 
For the interpretation of the solutions of the resonant swing spring we 
need some elementary facts about these ellipses which are the content 
of the following \medskip

\noindent \textbf{Lemma 1.} \emph{The center of $E$ is at the origin 
and its major axis makes an angle
\begin{equation}
\vartheta  = \onehalf {\tan }^{-1} \frac{2(xy+p_xp_y)}
{p^2_x+x^2- p^2_y -y^2 }
\end{equation}
with the $x$-axis, assuming that $p_x^2+x^2 > p_y^2+y^2$.
Otherwise, $\vartheta$ is the angle between the major axis and the $y$-axis.
In the coordinates $(\xi ,\eta , \zeta , p_{\xi },
p_{\eta },p_{\zeta })$ on $(T^{\ast }{\R }^3, \widehat{\omega })$ this 
angle is
\begin{equation}
\widetilde{\vartheta } =  {\Psi }^{\ast }\vartheta \, = \,
\onehalf {\tan }^{-1}\frac{\xi \eta + p_{\xi}p_{\eta }}
{\xi p_{\eta }-\eta p_{\xi}}
=  \,\frac{\pi}{2} - \onehalf {\tan }^{-1}\frac{\xi }{p_{\xi }}
+\onehalf {\tan }^{-1}\frac{{\eta }}{p_\eta }.
\end{equation}
The area of the ellipse $E$ is given by $\pi (xp_y-yp_x)^2 = \pi L^2$ 
and its eccentricity is
\begin{equation}
\mathrm{e}(E) = 2 \left( \kappa + \kappa^{-1} \right)^{-1},
   \qquad \kappa^4  = \frac{2j_1 - \rho_6}{2j_2 - \rho_6}  \,.
\end{equation}
}
\noindent \textbf{Proof.} Recall that (see \cite{cushman-bates})
\begin{equation}
\begin{array}{rlcrl} 
{\tau }_1 & = x^2 + p^2_x & \quad & {\tau }_3 & = xy +p_xp_y \\
\rowspace {\tau }_2 & = y^2 +p^2_y & \quad & {\tau }_4 & = xp_y-yp_x.
\end{array}
\label{eq-prfone}
\end{equation}
are integrals of the $2$-dimensional harmonic oscillator vector field 
\begin{equation}
X_{H^0_{xy}} = p_x\, \frac{\partial }{\partial x} + 
p_y\, \frac{\partial }y - x\, \frac{\partial }{\partial p_x} -
y\, \frac{\partial }{\partial p_y}.
\label{eq-prftwo}
\end{equation}
Moreover they satisfy the relation 
\begin{equation}
{\tau }^2_3 +{\tau }^2_4 = {\tau }_1{\tau }_2, \, \, \, 
{\tau }_1 \ge 0, \, {\tau }_2 \ge 0.
\label{eq-prfthree}
\end{equation}
The projection of a motion of the harmonic oscillator with initial 
condition $(x^0,y^0,p^0_x,p^0_y)$ and positive energy onto 
the $(x,y)$-plane is the curve
\begin{equation}
\gamma : t \mapsto \gamma (t) = (x(t), y(t)) \, = \, 
(x\cot t +p_x \sin t, y \cos t + p_y \sin t). 
\label{eq-prffour}
\end{equation}
The initial condition determines the values of the integrals ${\tau }_i$, 
say ${\tau }^0_i$. Then 
\begin{displaymath}
({\tau }^0_1 -{x(t)}^2)({\tau }^0_2-{y(t)}^2) = (p_x(t)p_y(t))^2 \, = \, 
({\tau }^0_3 - x(t)y(t))^2, 
\end{displaymath}
which after some simplification gives 
\[
{\tau }^0_2\, {x(t)}^2 -2{\tau }^0_3\, x(t)y(t) + {\tau }^0_1\, {y(t)}^2 = 
{\tau }^0_1{\tau }^0_2 - ({\tau }^0_3)^2 \, = \, ({\tau }^0_4)^2. 
\]
Since ${\tau }^0_1 +{\tau }^0_2 = 2H^0_{xy} >0$ and 
${\tau }^0_1{\tau }^0_2 - ({\tau }^0_3)^2 \ge 0$, the quadratic form 
\begin{equation}
(x,y)\mathcal{Q}\, \mbox{{\footnotesize $\left( \begin{array}{c} 
x \\ y \end{array} \right) $}} \, = \, 
(x,y) \mbox{{\footnotesize $\left( \begin{array}{rr} 
{\tau }^0_2 & - {\tau }^0_3 \\ -{\tau }^0_3 & {\tau }^0_1 \end{array} 
\right) $}} \, \mbox{{\footnotesize $\left( \begin{array}{c} 
x \\ y \end{array} \right) $}} \, = \, ({\tau }^0_4)^2
\label{eq-prfsix}
\end{equation}
is positive definite. Hence the image of the curve $\gamma $ is an 
ellipse $E$. 

Changing coordinates by {\tiny $ \left( \begin{array}{c} 
x \\ y \end{array} \right) = \left( \begin{array}{cr} \cos \theta & 
-\sin \theta \\
\sin \theta & \cos \theta \end{array} \right) \left( \begin{array}{c} 
\xi \\ \eta  \end{array} \right) $} transforms the quadratic 
form $\mathcal{Q}$ 
(\ref{eq-prfsix}) into the quadratic form 
\begin{equation}
(\xi , \eta ) \mbox{{\footnotesize $ \left( 
\begin{array}{cc} 
c^2\, {\tau }^0_2 + s^2\, {\tau }^0_1 -2cs \, {\tau }^0_3 & 
cs ({\tau }^0_1-{\tau }^0_2) -(c^2-s^2){\tau }^0_3 \\ 
cs ({\tau }^0_1 - {\tau }^0_2) -(c^2-s^2){\tau }^0_3 & 
s^2\, {\tau }^0_2 +c^2\, {\tau }^0_1 + 2cs \, {\tau }^0_3 
\end{array} \right) $ }} 
\, \mbox{{\footnotesize $ \left( \begin{array}{c} 
\xi \\ \eta \end{array} \right) $}}, 
\label{eq-prfseven}
\end{equation}
where $c = \cos \theta $ and $s = \sin \theta $. Suppose that ${\tau }^0_1 > 
{\tau }^0_2$. Choose $\theta $ so that 
\begin{equation}
\theta = \onehalf {\tan }^{-1} \frac{2{\tau }^0_3}{{\tau }^0_1-{\tau }^0_2}.
\label{eq-prfeight}
\end{equation}
Then the quadratic form (\ref{eq-prfseven}) becomes 
$
{\lambda }_{-}\, {\xi }^2 + {\lambda }_{+}\, {\eta }^2 = ({\tau }^0_4)^2, 
$
where ${\lambda }_{\pm }$ are eigenvalues of $\mathcal{Q}$, that is, 
\begin{equation}
{\lambda }_{\pm } = \onehalf \left[ ({\tau }^0_1 + {\tau }^0_2) \pm 
\sqrt{({\tau }^0_1+{\tau }^0_2)^2 -4({\tau }^0_4)^2} \right] > 0.
\end{equation}
Since ${\lambda }_{+} > {\lambda }_{-} > 0$, a piece of the $\xi $-axis is 
the major axis of $E$ and $\theta $ (\ref{eq-prfeight}) is the angle 
between the major axis and the $x$-axis. Note that the eccentricity $e$ of the 
ellipse $E$ is $e^2 = 1 - \frac{{\lambda }_{-}}{{\lambda }_{+}}$. \medskip 

We now compute $\kappa $. By definition 
\begin{eqnarray*}
{\kappa }^4 & = & \frac{2j_1 -{\rho }_6}{2j_2 -{\rho }_6} \, = \, 
\frac{{\rho }_1}{{\rho }_2} \, =\,
\frac{{\xi }^2 +p^2_{\xi }}{{\eta }^2 +p^2_{\eta }} \\
& = & \frac{x^2+p^2_x+y^2+p^2_y-2(xp_y-yp_x)}{x^2+p^2_x+y^2+p^2_y+2(xp_y-yp_x)} 
\, = \, \frac{{\tau }^0_1 +{\tau }^0_2 - 2{\tau }^0_4}
{{\tau }^0_1 +{\tau }^0_2 + 2{\tau }^0_4} \\ 
& = & \frac{({\tau }^0_1 +{\tau }^0_2)^2 - 2({\tau }^0_4)^2}
{({\tau }^0_1 +{\tau }^0_2 + 2{\tau }^0_4)^2} .
\end{eqnarray*}
So 
\begin{eqnarray*}
{\kappa }^2 & = & \frac{\sqrt{({\tau }^0_1 +{\tau }^0_2)^2 - 2({\tau }^0_4)^2}}
{{\tau }^0_1 +{\tau }^0_2 + 2{\tau }^0_4} \\
& = & \frac{{\lambda }_{+}-{\lambda }_{-}}{ {\lambda }_{+}+{\lambda }_{-} 
+2\sqrt{{\lambda }_{+}{\lambda }_{-}} }, \quad \parbox[t]{2.5in}{since the 
characteristic polynomial of $\mathcal{Q}$ is ${\lambda }^2 - ({\tau }^0_1 
+{\tau }^0_2)\lambda + ({\tau }^0_4)^2$} \\
& = & \frac{1 - \sqrt{\frac{{\lambda }_{-}}{{\lambda }_{+}}}}
{1 + \sqrt{\frac{{\lambda }_{-}}{{\lambda }_{+}}}}.
\end{eqnarray*}
Therefore 
\begin{displaymath}
1 - e^2 = \frac{{\lambda }_{+}}{{\lambda }_{-}} \, = \, 
\frac{(1-{\kappa }^2)^2}{(1+{\kappa }^2)^2},
\end{displaymath}
which implies that $e^2 = \frac{4{\kappa }^2}{(1+{\kappa }^2)^2}$, that is, 
$e = \frac{2\kappa }{1+{\kappa }^2} = \frac{2}{\kappa +{\kappa }^{-1}}$. 
\hfill $\square $ \medskip  

\begin{figure}
\centerline{\includegraphics[width=5cm]{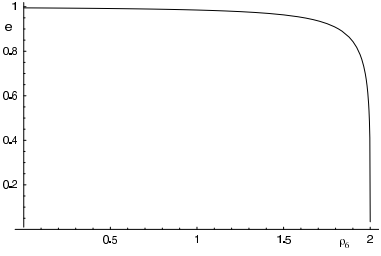}}
\caption{The eccentricity $e(\rho_6)$ for $j_1 = 1$, $j_2 = 1.5$. The 
graph is even closer to one when $j_1$ and $j_2$ are closer to each other.  
\label{FigEccent}}
\end{figure}

For small but non-zero angular momentum $l$ the motion goes from 
a small circle to a very eccentric ellipse, see Figure~\ref{FigEccent}.
When $l=0$ then $j_1=j_2$, hence $\kappa = 1$ and e = 1: 
the motion takes plane on a line, always. However, when 
$j_1 \not = j_2$ then the eccentricity will move between 
$0$ and $1 - 2l/j$. $E$ is close to $1$ for most values of $\rho_6$, 
and $\vartheta $ gives the angle of the swing plane. 
Only when $\rho_6$ is near its maximum the eccentricity becomes small. 
For the full Hamiltonian $\overline{H}$ of the swing spring, 
the angle $\vartheta$
is no longer constant, but instead is slowly varying. In this situation,
${\vartheta }$ is the instantaneous angle of the swing plane. 
Similarly $\rho_6$ is changing in time, so $e$ becomes the instantaneous 
eccentricity.
Even in the full dynamics the instantaneous ellipse $E$ has constant
area $\pi L^2$, because $L$ is the conserved angular momentum.
In general the eccentricity of $E$ changes during the motion of the
swing spring. However, since $\mathrm{e}(E)$ only depends on $\rho_6$, 
we see that it is constant for the critical 2-tori corresponding to 
a stable equilibrium point of the fully reduced vector field 
$X_{H_{j_1,j_2}}$. The swing angle $\vartheta $ should not be confused with
the polar angle of polar coordinates in the $(x,y)$-plane, even though
$\vartheta $ is canonically conjugate to $L$. As we will see
the angle $\vartheta$ commutes with $H_0$, which is not true
for the polar angle.\medskip

We now construct the action-angle variables corresponding to the
$T^2$ symmetry group.
In section $3$, we have found a Hamiltonian $2$-torus action 
(\ref{T2action}) on $T^{\ast }{\R}^3$ with momentum map $J = (J^1,J^2)$ 
(\ref{eq-s3thirtyseven}). For a regular value $(j_1,j_2)$ of $J$, the 
two-torus action on the level set
$J^{-1}(j_1,j_2) = (J^1)^{-1}(j_1)\cap (J^2)^{-1}(j_2)$ is free. Hence 
we obtain a smooth $2$-torus bundle ${\pi }_{j_1,j_2}: J^{-1}(j_1,j_2) 
\rightarrow
P_{j_1,j_2}$ over the fully reduced space $P_{j_1,j_2}$. We would like to find
action-angle coordinates for the $2$-torus fiber 
$F_p = {\pi }^{-1}_{j_1,j_2}(p)$
over the point $p \in P_{j_1,j_2}$. The Hamiltonian vector fields
$X_{J^1}$ and $X_{J^2}$ on $(T^{\ast}{\R }^3, \widehat{\omega })$ leave
each $2$-torus $F_p$ invariant and have only periodic orbits of
period $2\pi $ on $F_p$.
The same is therefore true for ${\widehat{H}}_0 = J^1 + J^2$ and 
$\widehat{L} = J^2 - J^1$.
Since we are interested in the angle $\theta$, we use ${\widehat{H}}_0$ and 
$\widehat{L}$ as action variables instead of $J^1$ and $J^2$.
We denote the values of these actions by $j$ and 
$l$, respectively. This leads to the following \medskip

\noindent \textbf{Lemma 2.} \emph{For all regular values of $J$ 
(\ref{eq-s3thirtyseven}) the variables $(I_1,\theta_1,I_2,\theta_2)$
given by $I_1 = {\widehat{H}}_0$, $I_2 = \widehat{L}$, (see
(\ref{eq-s2seventeen}) and (\ref{eq-s2eighteen})),  
\begin{equation}
    \theta_1 = \onehalf \tan^{-1} \frac{\zeta}{p_\zeta}, \qquad
    \theta_2 = \onehalf \tan^{-1} \frac{\eta}{p_\eta}
             - \onehalf \tan^{-1} \frac{\xi}{p_\xi}
\end{equation}
and ${\rho }_4, {\rho }_5, {\rho }_6$ on $P_{j_1,j_2}$ are a 
coordinate system for an 
open subset of phase space $T^{\ast }{\R }^3$. In these coordinates, the 
Hamiltonian of the resonant swing spring is
\begin{equation} 
\label{Hfin}
H = I_1 + \lambda \, \rho_4\,.
\end{equation}
Moreover, the following Poisson bracket relations hold.
\begin{subequations}
\begin{align}
\{ \theta_1, I_1 \} & = 1, &
\{ \theta_2 , I_2 \} & = 1, \\
\{ \theta_1, \rho_4 \} & = \onehalf \frac{\rho_4}{\rho_6},  &
\{ \theta_1, \rho_5 \} & = \onehalf \frac{\rho_5}{\rho_6}, \\
\{ \theta_2, \rho_4 \} & = -\frac{\rho_4 I_2}{\rho_1\rho_2}, &
\{ \theta_2, \rho_5 \} & = -\frac{\rho_5 I_2}{\rho_1\rho_2}, \\
\{ \theta_1, \rho_6 \} &= 1, &
\{ {\rho }_4, {\rho }_5 \} & = \rho_6(\rho_1+\rho_2) - \rho_1 \rho_2 , \\
\{ {\rho }_5, {\rho }_6\} & = 2 {\rho }_4, &
\{ {\rho }_4, {\rho }_6 \} & = -2 {\rho }_5,
\end{align}
\end{subequations}
where $\rho_1 = I_1-I_2-\rho_6$ and $\rho_2 = I_1+ I_2- \rho_6$.
All other brackets vanish.
} \medskip

\noindent \textbf{Proof.}
This last assertion can be verified by direct calculation using the old 
variables $(\xi ,\eta , \zeta , p_{\xi }, p_{\eta }, p_{\zeta })$. For example
\begin{eqnarray*}
\{ {\theta }_2, I_1 \} & = & \mbox{{\footnotesize $\left( p_{\zeta }
\frac{\partial }{\partial \zeta }
- \zeta \frac{\partial }{\partial p_{\zeta }} \right) $}}
{\tan }^{-1} \frac{p_{\zeta }}{\zeta } \, = \, 1. \\
\{ {\theta }_1, I_2 \} & = & 0 \\
\{ {\theta }_2 , I_1 \} & = & \mbox{{\footnotesize $ \left( + p_{\xi } 
\frac{\partial }{\partial \xi }
+ p_{\eta } \frac{\partial }{\partial \eta } -\xi  \frac{\partial }
{\partial p_{\xi }}
-\eta \frac{\partial }{\partial p_{\eta }} \right) $}} {\theta }_2
\, = \, \onehalf - \onehalf \, = \, 0 \\
\{ {\theta }_2, I_2 \} & = & \mbox{{\footnotesize $\left( - p_{\xi } 
\frac{\partial }{\partial \xi }
+ p_{\eta } \frac{\partial }{\partial \eta } + \xi  \frac{\partial }
{\partial p_{\xi }}
- \eta \frac{\partial }{\partial p_{\eta }}
\right) $}} {\theta }_2 \, = \, 1 \\
\{ {\theta }_1, \rho_4\} &=& \frac{(\xi\eta - p_\xi p_\eta) p_\zeta - 
(\xi p_\eta + \eta p_\xi) \zeta}{2(p_\zeta^2+\zeta^2)} = 
\frac{\rho_4}{2\rho_6} \\
\{ {\theta }_2, \rho_4\} &=& 
\frac{(p_\xi^2+\xi^2-p_\eta^2-\eta^2)((\xi\eta - p_\xi p_\eta) 
p_\zeta - (\xi p_\eta + \eta p_\xi) \zeta)}{2(p_\xi^2+\xi^2)(p_\eta^2+\eta^2)} = 
-\frac{{\rho }_4I_2}{{\rho }_1{\rho }_2} ,
\end{eqnarray*}
and similarly for the other brackets. That all brackets between the 
new variables can be expressed in terms of the $\rho_i$
follows from the fact that 
\par 1) $I_1$ and $I_2$ are actions; 
\par 2) the flow of the Hamiltonian vector fields of $I_1$ and $I_2$ leaves 
$\rho_i$ invariant; 
\par 3) the actions $I_1$ and $I_2$ and the angles $\theta_1$, $\theta_2$
form a symplectic coordinate system in the two-torus fibre of the 
bundle ${\pi }_{j_1,j_2}$. \hfill $\square $ \medskip

Lemma 2 allows us to give a description of
the motion of the resonant swing spring in phase space 
in such a way that the fully reduced system is 
an invariant subsystem that is driving the dynamics in the fibres. 
Using the fact that $\lambda \rho_4 = h-j$, the equations 
of motion for the angles in the fibre are
\begin{subequations}
\begin{align} 
\label{dottheta1}
\dot\theta_1 &= \{\theta_1,H\} = 1 + \frac{h-j}{2\rho_6}, \\
\dot\theta_2 &= \{\theta_2,H\} = \frac{(h-j)l}{(\rho_6-j-l)
(\rho_6-j+l)} \,.
\label{dottheta2}
\end{align}
\end{subequations}
They are driven by the solution of the the 
second order differential equation 
\begin{equation} 
\label{ddotrho6}
\ddot \rho_6 = 2 \lambda^2 \, P_3'(\rho_6)
     \, = \, 2\lambda^2 (2\rho_6(j-\rho_6) -(\rho_6-j-l)(\rho_6-j+l))
\end{equation}
Equations (\ref{dottheta2}) and (\ref{ddotrho6}) can be found in \cite{HL}.
We rederived them because in \cite{HL} they were derived 
under the hypothesis of ``pattern evocation in shape space'', which might have
been an approximation. From our derivation we see that \emph{no} approximation 
is involved and that the angle of the swing plane is 
simply an angle of an action-angle coordinate system.

\section{Actions and Rotation Numbers}

We are interested in the change of $\theta_2$ over one period of the
motion of $\rho_6$. First we directly derive an integral for
this change $\Delta \theta_2$ by changing the time parametrization
in (\ref{dottheta2}). Then we will show that $\Delta\theta_2$
is a rotation number.

The integral curve in the fully reduced system is given by the $h$-level 
set of the
fully reduced Hamiltonian $H_{j_1,j_2}$ (\ref{Hfin}) on the
fully reduced space $P_{j_1,j_2}$ (\ref{Gdef}).
Hence the motion takes place on
a family of real affine elliptic curves ${\mathcal{E}}_{h,j_1,j_2}$
defined by
\begin{equation}
{\rho }^2_5 =  Q(\rho_6) = -\frac{1}{\lambda^2}(h-j)^2 + \rho_6 
(\rho_6-j-l)(\rho_6-j+l)  \,,
\label{rho5issQ}
\end{equation}
when $0 \le {\rho }_6 \le \min (2j_1,2j_2)$. Equation (\ref{rho5issQ})
is obtained by eliminating ${\rho }_4$ from
(\ref{eq-s3thirtyseven}) using $h = j_1+j_2 + \lambda {\rho }_4$ (which
defines $H^{-1}_{j_1,j_2}(h))$. This exactly reproduces the polynomial
$Q$ already defined in (\ref{Qdef}). Therefore also the analysis of the
double roots done for the energy momentum map also applies here.
The equation (\ref{ddotrho6}) for $\rho_6$ can be 
integrated once and we find
\begin{equation} 
\label{dotrho6}
   \dot \rho_6 = 2\lambda \rho_5 =   2\lambda \sqrt{ Q(\rho_6) }  \,.
\end{equation}
Note that this is one of the fully reduced equations of motion 
(\ref{eq-s3fortystar}).
When $(h,j_1,j_2)$ is a regular value of
the energy momentum mapping $\mathcal{EM}$ (\ref{eq-s2twentyone}) of
the swing spring (which we henceforth assume), then
the ${\mathcal{E}}_{h,j_1,j_2}$ is smooth. When $h=j$ the
polynomial $P_3(\rho_6) = {\rho }_6(j+l-{\rho }_6)(j-l-{\rho }_6)$ is 
nonnegative on
$[0, \min(2j_1,2j_2)]$. Therefore the polynomial $Q$ has three 
\emph{distinct} real roots
\begin{displaymath}
0 \le {\rho }^{-}_6 < {\rho }^{+}_6 \le \min (2j_1,2j_2) < {\rho }^0_6.
\end{displaymath}
The motion of the fully reduced system on $H^{-1}(h)$ takes place when
${\rho }_6$ lies in $[{\rho }^{-}_6, {\rho }^{+}_6]$. Since 
$H^{-1}_{j_1,j_2}(h)$
is diffeomorphic to a circle, when the fully reduced motion runs through a
period the time parameter ${\rho }_6$ traverses the interval 
$[{\rho }^{-}_6,{\rho }^{+}_6]$ forward and backward once.
Accordingly the period of the driving motion is given by
\begin{equation} \label{Tper}
T = 2 \int_{\rho_6^-}^{\rho_6^+} \frac{\dee \rho_6}{2\lambda \rho_5}
\end{equation}

To find the change of $\theta_2$ over this period, we 
introduce in (\ref{dottheta1}) 
and (\ref{dottheta2}) a new time scale ${\rho }_6$ defined by (\ref{dotrho6}). 
We obtain
\begin{subequations}
\begin{eqnarray}
\dee {\theta }_1 & = &
\left( 1 + \frac{h-j}{2\rho_6}\right)  \, \frac{\dee {\rho }_6} 
{2\lambda{\rho }_5}
\label{eq-s6fiftyeight} \\
\dee {\theta }_2 & = & \frac{(h-j)l}{(j+l -{\rho }_6)(j-l-{\rho }_6)} \,
\frac{\dee {\rho }_6}{2\lambda {\rho }_5}
\label{dtheta2drho6}
\end{eqnarray}
\end{subequations}
which are differential forms on ${\cal E}_{h,j_1,j_2}$. 
Therefore the change in the
swing angle during a period of the motion of the swing spring is
\begin{eqnarray}
\Delta {\theta }_2(h,j,l) & = & 2\int^{{\rho}^{+}_6}_{{\rho }^{-}_6} 
\dee {\theta }_2 \nonumber \\
& = & \frac{(h-j)l}{\lambda}
\int^{{\rho }^{+}_6}_{{\rho }^{-}_6} \frac{1}{(j+l- {\rho }_6)
(j-l-{\rho }_6)}\, \frac{\dee {\rho }_6}{\sqrt{Q({\rho}_6)}}.
\label{eq-s6sixtydot}
\end{eqnarray}

To see that $\Delta {\theta }_2$ is in fact a rotation number of the integrable 
system $(\widehat{H}, J^1,J^2, T^{\ast }{\R }^3, \widehat{\omega })$, 
we have to compute the third action. First we find 
the angle $\theta_3 $ conjugate to ${\rho }_6$ on $P_{j_1,j_2}\setminus 
\{ ({\rho }^{\pm }_6,0) \}$. A calculation shows that
\begin{equation}
{\theta }_3 = \onehalf  \tan^{-1} \frac{\rho_5}{\rho_4} .
\end{equation}
does the job, because $\{ {\theta }_3, {\rho }_6 \} =1$. Therefore, 
we can define the third action $I_3$ as the integral of the 
canonical one form ${\rho }_6\, \dee {\theta }_3 $ over ${\cal E}_{h,j_1,j_2}$, 
namely, 
\begin{equation}
  I_3(h,j,l) = \frac{1}{2\pi} \oint \rho_6 \dee {\theta }_3 
    = \frac{h-j}{8\pi \lambda } \int_{\rho_6^-}^{\rho_6^+}
          \left( 3 - \frac{2j_1}{\rho_1} - \frac{2j_2}{\rho_2} \right) 
\frac{ \dee \rho_6}{\rho_5} . 
\label{thirdaction}
\end{equation}
The last equality above is verified as follows. Computing the 
derivative of ${\theta }_3$ along the integral curves of the 
fully reduced vector field $X_{j_1,j_2}$ and using (\ref{eq-s3fortystar})
gives
\begin{equation}
\frac{\dee {\theta }_3}{\dee t} = \onehalf \frac{{\rho }_4}{{\rho }^2_4 
+{\rho }^2_5} \, \frac{\dee {\rho }_5}{\dee t} \, = \, 
\frac{h-j}{2\lambda } \, \frac{\lambda }{{\rho }_1 {\rho }_2 {\rho }_6} \, 
P'_3({\rho }_6).
\end{equation}
So 
\begin{equation}
\frac{\dee {\theta }_3}{\dee {\rho }_6} = \frac{h-j}{4\lambda } 
\frac{({\rho }_1 {\rho }_2 - {\rho }_2 {\rho }_6 - {\rho }_1{\rho }_6)}
{{\rho }_1{\rho }_2 {\rho }_5 {\rho }_6}.
\end{equation}
Substituting ${\rho }_6 = 2j_1- {\rho }_1$ and ${\rho }_6 = 2j_2 -{\rho }_2$ 
gives the equality (\ref{thirdaction}).

The integral (\ref{thirdaction}) implicitly defines the 
Hamiltonian $H$ as a function of the three actions $I_1$, $I_2$, $I_3$.
Hence we can obtain the frequencies of the conjugate angles ${\theta }_1$, 
${\theta }_2$, 
and ${\theta }_3$ by implicit differentiation. In particular we have
\begin{equation}
\dee I_3 = \frac{\partial I_3}{\partial h} \, \dee H + \frac{\partial 
I_3}{\partial j} \, \dee I_1 + \frac{\partial I_3}{\partial l} \, \dee I_2 .
\end{equation}
Rewritten this gives
\begin{equation}
\begin{aligned}
\dee H & =  -{\left( \frac{\partial I_3}{\partial h } \right)}^{-1}
\frac{\partial I_3}
{\partial j} \, \dee I_1 - {\left( \frac{\partial I_3}{\partial h } \right)}^{-1}
\frac{\partial I_3}{\partial l} \, \dee I_2 + 
{\left( \frac{\partial I_3}{\partial h } \right)}^{-1}\, \dee I_3 \\
  & =   \frac{\partial H}{\partial I_1}\, \dee I_1 + 
\frac{\partial H}{\partial I_2}\, \dee I_2 + 
\frac{\partial H}{\partial I_3}\, \dee I_3.
\end{aligned}
\end{equation}
Hence 
\begin{equation}
{\dot{\theta}}_3 =  \frac{\partial H}{\partial I_3} =  
\left( \frac{\partial I_3}{\partial h} \right)^{-1}.
\label{eq-s6sixtynewdot}
\end{equation}
The above expression yields the period $2\pi \, \frac{\partial I_3}{\partial h}$, 
which is the same as that given in (\ref{Tper}). To see that the integral 
(\ref{Tper}) and the one obtained from (\ref{eq-s6sixtynewdot}) 
give the same result, 
one has to add a total differential to the integrand of the loop integral 
(\ref{Tper}), 
which does not contribute to the period. More precisely the identity
\begin{equation}
\frac{\dee \rho_6}{ \rho_5} =   
\frac{\partial }{\partial h} \left( \frac{h-j}{2 }\left( 3 - \frac{2j_1}
{{\rho }_1} - \frac{2j_2}{{\rho }_2} \right) 
\frac{ \dee {\rho }_6 }{ \rho_5 } \right) + \dee \left(  
\frac{\rho_6}{  \rho_5} \right) \,,
\end{equation}
holds on ${\mathcal{E}}_{h,j_1,j_2}$, where $\rho_5$ is a 
function of $\rho_6$ given by  (\ref{rho5issQ}). 
The frequencies of the other angles are similarly obtained as
\begin{equation}
{\dot{\theta }}_2 = \frac{\partial H}{\partial I_2} =  
-\frac{\partial I_3/\partial l}{\partial I_3/\partial h}
\end{equation}
and similarly for $\theta_1$. Hence we obtain the rotation numbers
\begin{equation}
   W_{23}  = \frac{{\dot{\theta }}_2}{{\dot{\theta }}_3}  =  
   -\frac{\partial I_3}{\partial l}, \, \, \mathrm{and} \, \, 
  W_{13} =  \frac{{\dot{\theta }}_1}{{\dot{\theta }}_3} =  
  -\frac{\partial I_3}
  {\partial j }\,.
\end{equation}
Now we show that $\Delta \theta_2 = -W_{23}$.
The equality holds because the differentials only differ by a total 
differential, which does not
contribute under the closed loop integral. By direct calculation one 
can check that
\begin{equation}
\dee \theta_2 = \frac{\partial }
{\partial l}\left( \frac{h-j}{4\lambda }\left( 3 - \frac{2j_1}{{\rho }_1} 
- \frac{2j_2}{{\rho }_2} \right) \frac{ \dee {\rho }_6 }{ \rho_5 } \right) 
+  \dee \left( \rho_6 \frac{\dee \theta_2}{\dee \rho_6} \right) \,,
\end{equation}
where $\dee \theta_2$ is given by (\ref{dtheta2drho6}).
\begin{figure}
\centerline{\includegraphics[width=7cm]{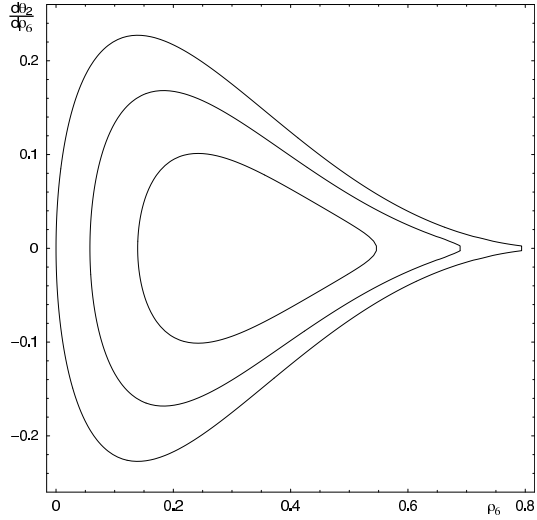}}
\caption{Contour plot of $\chi$ for $j = 1$, $l = 0.15$. 
The contours $0, -0.05, -0.1$ are shown.
\label{FigChi}}
\end{figure}
The first derivation of an integral for the stepwise precession of 
the swing plane 
is simpler than the present one. But the latter derivation shows 
that the stepwise precession of the swing plane is one of the two 
rotation numbers
of the invariant tori of the integrable approximation.

In order to understand the solution curves of the differential equation
for $\dee\theta_2/\dee\rho_6$ (\ref{dtheta2drho6}) we can plot its 
solution curves.
This is easily done by finding a function of $\theta_2$ and $\rho_6$ that is
constant on the solutions.
Such a function can be defined as
\begin{equation}
\chi(x,y) = \frac{y^2}{((x-j)^2 - l^2)^2} - x((x-j)^2-l^2) \,
\end{equation}
where  $\chi(\rho_6,\dee\theta_2/\dee\rho_6) = 
-(h-j)^2/\lambda^2$.
A contour plot of $\chi$ is shown in Figure~\ref{FigChi}. 
On the right part of figure the $\rho_6$ is large, so that the 
eccentricity is small. On the left side $\rho_6$ is small, hence the 
eccentricity is close to 1, see Figure~\ref{FigEccent}.

\section{Analysis of the swing angle}

In this section we examine more closely our formula (\ref{eq-s6sixtydot}) 
for the swing angle of the swing spring. We prove the following \medskip 

\noindent {\bf Proposition.} \emph{The rotation number describing the
angle of stepwise precession for the
integrable approximation of the resonant swing spring is given by
\begin{equation}
\Delta \theta_2 = \tan^{-1} \frac{l\sqrt{j}}{(h-j)/\lambda} + O\left 
(\sqrt{l^2 + ((h-j)/\lambda)^2/j}/j\right  )\,.
\label{eq-s10one}
\end{equation}
}
Equation (\ref{eq-s10one}) shows that
the swing angle $\Delta {\theta }_2$ is a multivalued function of the 
parameters $h$, $j_1$ and $j_2$. From this one can read off that
the Liouville integrable system $(\widehat{H}, \widehat{L}, {\widehat{H}}_0,
T^{\ast }{\R }^3, \widehat{\omega })$ describing the swing spring has
monodromy. \medskip

\noindent {\bf Proof.}
In order to simplify our analytical study of (\ref{eq-s6sixtydot}) we remove
an unneeded parameter by the rescaling
\begin{equation}
{\rho }_6 = j \, \tilde{z}, \quad l \, = \, j \, \widetilde{b} \quad 
\mathrm{and}
\quad  \widetilde{a}j^{3/2} \, = \, \frac{h-j}{\lambda }.
\end{equation}
Equation (\ref{eq-s6sixtydot}) then becomes
\begin{equation}
\Delta {\theta }_2  = \widetilde{a}\widetilde{b} \,
\int^{\tilde z^{+}}_{\tilde z^{-}} \frac{1}{(1-\tilde z)^2 - 
{\widetilde{b}}^2} \,
\frac{\dee \tilde z}{\sqrt{\tilde z((1-\tilde z)^2-{\widetilde{b}}^2) 
- {\widetilde{a}}^2}} .
\label{eq-s7sixtyone}
\end{equation}
Here $0 \le \tilde z^{-} < \tilde z^{+} \le 1 < \tilde z^0$ are
distinct nonnegative roots of the polynomial
\begin{displaymath}
\tilde z((1-\tilde z^2)-{\widetilde{b}}^2) - {\widetilde{a}}^2.
\end{displaymath}
\par
Next we restrict the parameters $\widetilde{a}$ and $\widetilde{b}$ in
(\ref{eq-s7sixtyone}) to lie on a line through the
origin in parameter space ${\R }^2$ and assume that they are both small.
In other words, we introduce a small parameter $\varepsilon $ such that
\begin{equation}
\widetilde{a} = \varepsilon \, a \quad \mathrm{and} \quad 
\widetilde{b} = \varepsilon \, b,
\label{eq-s7sixtyonestar}
\end{equation}
where $(a,b)$ are fixed in parameter space.

Now we find
the Taylor expansion of the roots ${\widetilde z}^{-}$, ${\widetilde z}^{+}$, 
and ${\widetilde{z}}^0$. A calculation gives
\begin{eqnarray}
{\widetilde{z}}^{-} & = & a^2\, {\varepsilon }^2 + (2a^4+a^2b^2) 
{\varepsilon }^4 +
{\mathrm{O}}({\varepsilon }^6)
\label{eq-s7sixtyfour} \nonumber \\
{\widetilde{z}}^{+} &=& 1 - \sqrt{a^2+b^2}\, {\varepsilon } -\onehalf a^2\, 
\varepsilon^2 + {\mathrm{O}}({\varepsilon }^3) 
\label{eq-s7sixtyfive} \\ \nonumber
{\widetilde{z}}^0 &=& 1 + \sqrt{a^2+b^2}\, {\varepsilon }  + \onehalf a^2\, 
\varepsilon^2 + {\mathrm{O}}({\varepsilon }^3) .
\label{eq-s7sixtyfivestar}
\end{eqnarray}

Because the root ${\widetilde{z}}^{+}$ and the pole at $1-\varepsilon b$
coalesc at $1$ as $\varepsilon \to 0$, we introduce a shifted, 
scaled, and inverted new variable $z$ by
\begin{equation}
\widetilde{z} = 1 + \varepsilon b\, / z\, .
\label{eq-s7sixtytwo}
\end{equation}
The inversion ensures that the new integration boundaries are finite in the
limit $\varepsilon \to 0$.
Then (\ref{eq-s7sixtyone}) becomes
\begin{equation}
\Delta {\theta }_2 = \int^{{z }^{+}}_{{z }^{-}} \frac{az^2}{z^2-1} \,
\frac{\dee z }{\sqrt{-z[ (a^2+b^2)z^3 -  b^2z+ {\varepsilon } b^3 (z^2-1)]}},
\label{eq-s7sixtythree}
\end{equation}

Factor the polynomial under the square root as
\[
   -z[ (a^2+b^2)z^3 -  b^2z+ \varepsilon b^3 (z^2-1)] = (z-z^-)(z^+-z) 
z (z - z^0) (a^2+b^2)
\]
and introduce a new integration variable $\phi$ by
\begin{equation}
     2 z = (z^++z^-) + (z^+-z^-) \cos\phi \,.
\end{equation}
The integral (\ref{eq-s7sixtythree}) becomes
\begin{equation} 
\label{theint}
   \Delta \theta_2 = \frac{1}{\sqrt{a^2+b^2}} \int_0^\pi f(z(\phi)) \dee \phi , 
\end{equation}
where
\begin{equation}
   f(z) = \frac{az^2}{(z^2-1)\sqrt{z(z-z^0)}} \,.
\end{equation}
Using (\ref{eq-s7sixtytwo}) we find that (\ref{eq-s7sixtyfour}) becomes
\begin{align}
z^{-}  & =  - \frac{b}{\sqrt{a^2+b^2}} \varepsilon  - 
\frac{a^2b}{a^2+b^2} \varepsilon  + {\mathrm{O}}({\varepsilon }^2) \nonumber \\
z^{+}  & =  -b {\varepsilon } + {\mathrm{O}}({\varepsilon }^3)  \\
z^{0}  & =  + \frac{b}{\sqrt{a^2+b^2}} \varepsilon  - 
\frac{a^2b}{a^2+b^2} \varepsilon  + {\mathrm{O}}({\varepsilon }^2) . \nonumber
\end{align}
Expanding the integrand of (\ref{theint}) up through terms of order
$\varepsilon $ gives
\begin{equation}
\frac{-2ab (\cos\phi-1)^2}{(4a^2 + (3-\cos\phi)(1+\cos\phi) 
b^2)\sqrt{(\cos\phi-1)(\cos\phi-3)}}
         +{\mathrm{O}}({\varepsilon }) .
\label{integr} 
\end{equation}
Note that the error term is uniformly bounded in the interval of integration. 
The main purpose of the above transformations
was to achieve this. Now the zero order 
contribution in (\ref{theint}) can be calculated. The substitution 
$\cos\phi = 1 + 2 \sin\psi$ removes the root and the 
second substitution $\cos\psi = x$ rationalizes (\ref{integr}). Hence
the integral becomes
\begin{equation}
\int_0^1 \frac{ab \dee x}{a^2 + b^2x^2} = \tan^{-1} \frac{b}{a}\,.
\end{equation}
Undoing the scaling gives the desired result. \hfill $\square$ \medskip

Using the above method, one can actually compute one more order. But it
turns out to be zero. At $O(\varepsilon^2)$ the integrand is
not uniformly bounded. So more sophisticated methods are needed
to obtain the first nonzero correction. Note that the error term is small 
when we are close to the thread in the
bifurcation diagram. Then $l$ and $h-j$ are close to zero. Of course 
$j$ itself is also small, since we must be close to the equilibrium, 
but $l$ and $h-j$ are considered to be much smaller.

The result means that for each fixed value of $j>0$, the swing angle
$\Delta {\theta }_2$ is a multivalued function of $l$ and $h$.
This is another proof of the fact that the system has monodromy.
The resonant swing spring provides an example in which the monodromy
can be easily observed.

\section*{Acknowledgements}

This work was supported 
by the EU network HPRN-CT-2000-0113 \textit{MASIE -- Mechanics
and Symmetry in Europe}. A first draft of this paper was 
written at University of Warwick during the Symposium on 
Geometric Mechanics and Symmetry 2002. The authors would like to 
thank the Maths Research Centre at Warwick for its hospitality.

\end{document}